\documentclass[amsfonts,nofootinbib,superscriptaddress,aps,twocolumn]{revtex4-1}

\usepackage{amsmath}
\usepackage{amssymb}
\usepackage{appendix}
\usepackage{amsbsy}
\usepackage{bm}
\usepackage{epsfig} 
\usepackage{color}
\usepackage{slashed}
\usepackage{graphics}
\usepackage{subfigure}
\usepackage{graphicx}
\usepackage{lipsum}

\begin{document}
\title{A bi-metric universe with matter }
 \author{Carlos Maldonado}
 \email{carlos.maldonados@usach.cl}
\affiliation{Universidad de Santiago de Chile (USACH),
Facultad de Ciencia,
Departamento de  F\'{\i}sica, Chile.}
\author{Fernando M\'endez}
 \email{fernando.mendez@usach.cl}
\affiliation{Universidad de Santiago de Chile (USACH), Facultad de Ciencia,
Departamento de F\'{\i}sica, Chile.}

\begin{abstract}
We analyze the early stage of evolution of a universe with two scale factors 
proposed in  \cite{Falomir1} when matter is present. The scale factors
describe two  causally disconnected patches of the universe interacting
 trough  a non-trivial Poisson bracket  structure in the momentum 
sector characterized by one parameter $\kappa$. 
We studied two  scenarios  in which one of the patches 
is always filled with relativistic matter while the other contains
relativistic matter in one case, and non-relativistic matter in the second case. 
By solving numerically the set of  equations governing the dynamics, 
we  found that the  energy content of one sector {\it drains} to the other and from here 
it is possible to constraint  the deformation parameter $\kappa$ 
by imposing that the decay of the  energy density happens, at most, at the   Big Bang Nucleosynthesis 
temperature {in order to return to the usual behavior of radiation}. 
The relation with Non Standard Cosmologies is also addressed. 
\end{abstract}

\maketitle

\section{Introduction}

Our present description of the universe rests on the cosmological 
principle --  the hypotheses  of  spatial homogeneity  and isotropy
at large scales -- described by a  Friedman-Lemaitre-Robertson-Walker (FLRW)  metric
\cite{Weinberg2008,Kolb:1990vq}. Observations of 
rotational curves of galaxies \cite{Zwicky:1933gu,Freeman:1970mx} (for a review see 
\cite{Bertone:2016nfn,Salucci:2018}) as well as  the  observed  
accelerated cosmic expansion \cite{Riess_1998,Perlmutter_1999}, made
necessary to complete the model with two extra hypothesis: the existence
of dark matter and dark energy (cosmological constant term $\Lambda$),
respectively. Thus, our present
model of the universe, according to observations \cite{Zyla:2020zbs}, 
contains a $68,3 \%$ of dark energy, $26.8\%$ of cold (non-relativistic) 
dark matter and $4.9\%$ of baryonic matter.

On the other hand,  possible traces of inhomogeneities\footnote{The possibility of formation of cosmic strings, monopoles or domain walls 
can not be discarded from a theoretical point of view \cite{Vilenkin,Sikivie,Lukas,Flanagan:2000}.} 
have been smoothed out during the exponentially accelerated period of expansion known as {\it inflation} 
\cite{Peebles:1999,Linde:2005ht}, a new hypothesis which also solves the   flatness and  horizon problems,
explains the origin of large-scale  structures in the universe  and restores  homogeneity 
inside the cosmological horizon.

In this regard, in a recent set of papers 
\cite{Falomir1,Falomir:2018ayx,Falomir_2020}  a model for a universe with two  metrics was considered. 
In such model, two regions (patches) causally disconnected after the inflation era,
are described with  metrics of  FLRW type with different scale factors for each patch,   and a
sort of interaction was introduced through a deformation of the Poisson bracket structure in the space of fields.  
It was shown that, in absence of matter, this sort of interaction emulates the presence of cosmological constant 
on each patch.

In the present paper we extend the previous model in order to incorporate
matter  assuming that matter evolves independently on each 
sector and it can be modelled as a barotropic perfect fluid.  We will show 
that this model can be understood as a sort of Non Standard Cosmology (NSC) 
\cite{Scherrer:1985,Hamdan:2018,Giudice:2001,Gelmini:2006,Maldonado:2019,Arias:2019,Bernal:2019}, 
for different values of the parameter controlling  the Poisson's bracket deformation.

In order to do that, in the next section we will show the main features
of the model with two metrics and the NSC scenario. Section III is devoted 
to the discussion of how to incorporate matter into the model. 
In section IV two cases will be addressed: 
a) one patch filled  with relativistic matter while the second one contains a non-relativistic fluid 
and b) both patches containing relativistic matter. In the final section we present the conclusions and 
discuss possible extensions of the model.

\section{The two-metric universe}
The model discussed in  \cite{Falomir1} (see also \cite{Falomir:2018ayx,Falomir_2020}) describes
two patches of the universe through scale factors $a(t)$ an $b(t)$ and a Hamiltonian
\begin{eqnarray}
\label{ham1}
H&=&\frac{NG}{2} \left[\frac{\pi_a^2}{a}+\frac{1}{G^2}\left(a\,k_a-\frac{\Lambda_a}{3}a^3\right)\right]+
\nonumber
\\
&&
\frac{NG}{2} \left[\frac{\pi_b^2}{b}+\frac{1}{G^2}\left(b\,k_b -\frac{\Lambda_b}{3}b^3\right)\right],
\\
&\equiv&H_a+H_b
\end{eqnarray}
where $\pi_a,\pi_b$ are the conjugate momenta of $a$ and $b$, respectively. Scale factors are chosen 
with canonical dimension $-1$ and then, momenta have dimension +1\footnote{The canonical dimensions of
fields are chosen in this way just because a convenience matter.}.  $N$ is an auxiliary field 
that guaranties the time reparametrization invariance. Patches $a$, $b$ have spatial curvature 
$k_a$, $k_b$ and cosmological constant $\Lambda_a, \Lambda_b$, respectively. 

The Poisson bracket structure, on the other hand, is defined through the
following relations
\begin{equation}
\label{pb}
\{a_\alpha,a_\beta\} =0, \quad \{a_\alpha,\pi_\beta\}=\delta_{\alpha\beta},\quad 
\{\pi_\alpha,\pi_\beta\}=\theta\,\epsilon_{\alpha\beta},
\end{equation}
with $\theta$ a constant parameter and index $\{\alpha,\beta\}\in\{a,b\}$. Scale
factors notation is $a_a=a,a_b=b$. It is convenient to redefine the parameter
$\theta$ as $\theta = \kappa\,G^{-1}$ with $\kappa$ a dimensionless parameter.

Equations of motion derived from Hamiltonian (\ref{ham1}) with Poisson brackets
(\ref{pb}) are 
\begin{eqnarray}
\label{vel}
\dot{a} &=& G\frac{\pi_a}{a},\quad \dot{b}=G\frac{\pi_b}{b},
\\
\label{piadot}
\dot{\pi}_a &=&G\frac{\pi_a^2}{2a^2}+\frac{1}{2G}\left(a^2\Lambda_a-k_a\right)+\kappa\frac{\pi_b}{b},
\\
\label{pibdot}
\dot{\pi}_b &=&G\frac{\pi_b^2}{2b^2}+\frac{1}{2G}\left(b^2\Lambda_b-k_b\right)+\kappa\frac{\pi_a}{a},
\\ \nonumber
\end{eqnarray}
while the constraint $\dot{\pi}_N =0$ reads
\begin{equation}
\label{const}
\frac{\pi_a^2}{a}+\frac{\pi_b^2}{b}+G^{-2}\left(a\,k_a+b\,k_b- \frac{\Lambda_a}{3}\,a^3
-\frac{\Lambda_b}{3}\,b^3\right)=0.
\end{equation}
Note that we have written the equations in the usual gauge $N=1$ (equivalently, we have redefined the time variable $dt'=N(t)dt$).

Equations  (\ref{vel}) to  (\ref{const}) can be recast as the following  set of second order differential equations
\begin{eqnarray}
\label{eom2a}
2a\ddot{a}+\dot{a}^2 &=&\Lambda_a\,a^2-k_a+2\kappa\dot{b},
\\
\label{eom2b}
2b\ddot{b}+\dot{b}^2 &=&\Lambda_b\,b^2-k_b-2\kappa\dot{a},
\\
\label{const2}
a\dot{a}^2+b\dot{b}^2&=&\frac{\Lambda_a}{3}\,a^3-k_a\,a+
\frac{\Lambda_b}{3}\,b^3-k_b.
\end{eqnarray}

The model presents several interesting properties, as for example, the existence of solutions containing
both, accelerated and decelerated periods, or the presence of an inflationary epoch in a patch with a
negligible cosmological constant (for example, for $\Lambda_a\ll\Lambda_b$). Note also that 
the equations are symmetric under the simultaneous change $a\rightarrow b$, $b\rightarrow a$ and $\kappa\rightarrow -\kappa$.

The effect of matter in the model, on the other hand, has not been explored and it is the main  
purpose of the present work to investigate this scenario. We will show that this two-metric model with matter 
have  similar features compared with the Non Standard Cosmologies (NSCs) scenarios.

Indeed, the study of  {the effects of different cosmological histories at  early stages of the universe, such as matter domination ($H\propto T^{3/2}$) \cite{Giudice:2001}, kination domination ($H\propto T^6$) \cite{Visinelli:2018} or even a  field with a general state of equation ($H\propto T^{3(\omega+1)/2}$) \cite{Maldonado:2019,Arias:2019,Bernal:2019}, where $H$ is the Hubble parameter, is 
a very active field of research.}

{A particular scenario relevant to the present work  considers  the introduction of a  field ($\phi$)  whose only effect 
is to modify  the expansion rate of the universe, making  it faster or slower (and also can decay into Standard Model 
particles). These kind of different cosmological histories are usually called Non-Standard Cosmologies (NSCs) and the only 
restriction  for this new field is to decay before the epoch of Big Bang Nucleosinthesys (BBN), in order not to be in 
conflict with  astrophysical measures \cite{Chung:1999,Kolb:2003}}.

{For a general NSC model, the evolution equations read
\begin{eqnarray}
\label{radNSC}
\dot{\rho_\text{SM}}+4H\rho_\text{SM} &=& \Gamma \rho_\phi
\\
\label{phiNSC}
\dot{\rho_\phi}+3(\omega+1)H\rho_\phi &=& -\Gamma\rho_\phi,
\end{eqnarray}
where $\rho_\text{SM}$ is the energy density of the Standard Model (SM) content,
$\rho_\phi$ is the energy density of the new field, 
$\omega$ is the constant for the barotropic fluid, 
$\Gamma$ the decay constant for the $\phi$ field and $H$ is the Hubble parameter.}

{The decay constant   $\Gamma$ can be expressed in terms of the re-heating temperature $T_{RH}$ (sometimes called $T_\text{end}$ depending on the behavior of the $\phi$ field), by demanding
\begin{equation}
\Gamma=\frac{\pi}{3}\sqrt{\frac{g_\star}{10}}T_{RH}^2,
\end{equation}}
that is, $\Gamma$ is equal to the value of the Huble parameter at the
time when universe is dominated by radiation again. 
Here, $g_*$ is the degrees of freedom of radiation that 
we will consider as a constant with  value $g_\star\approx 10$,
which corresponds to a temperature for $T_\text{RH}$ (or $T_\text{end}$) of $T=4\times 10^{-3}$ GeV. 
This value is imposed by the BBN epoch and corresponds to the lowest value of the temperature 
at which this new field must decay \cite{Kawasaki:2000,Hannestad:2004}.

\begin{figure}[h!]
    \centering
    \begin{subfigure}[\label{fig:rhoinflaton}]{
         \centering
         \includegraphics[width=0.46\linewidth]{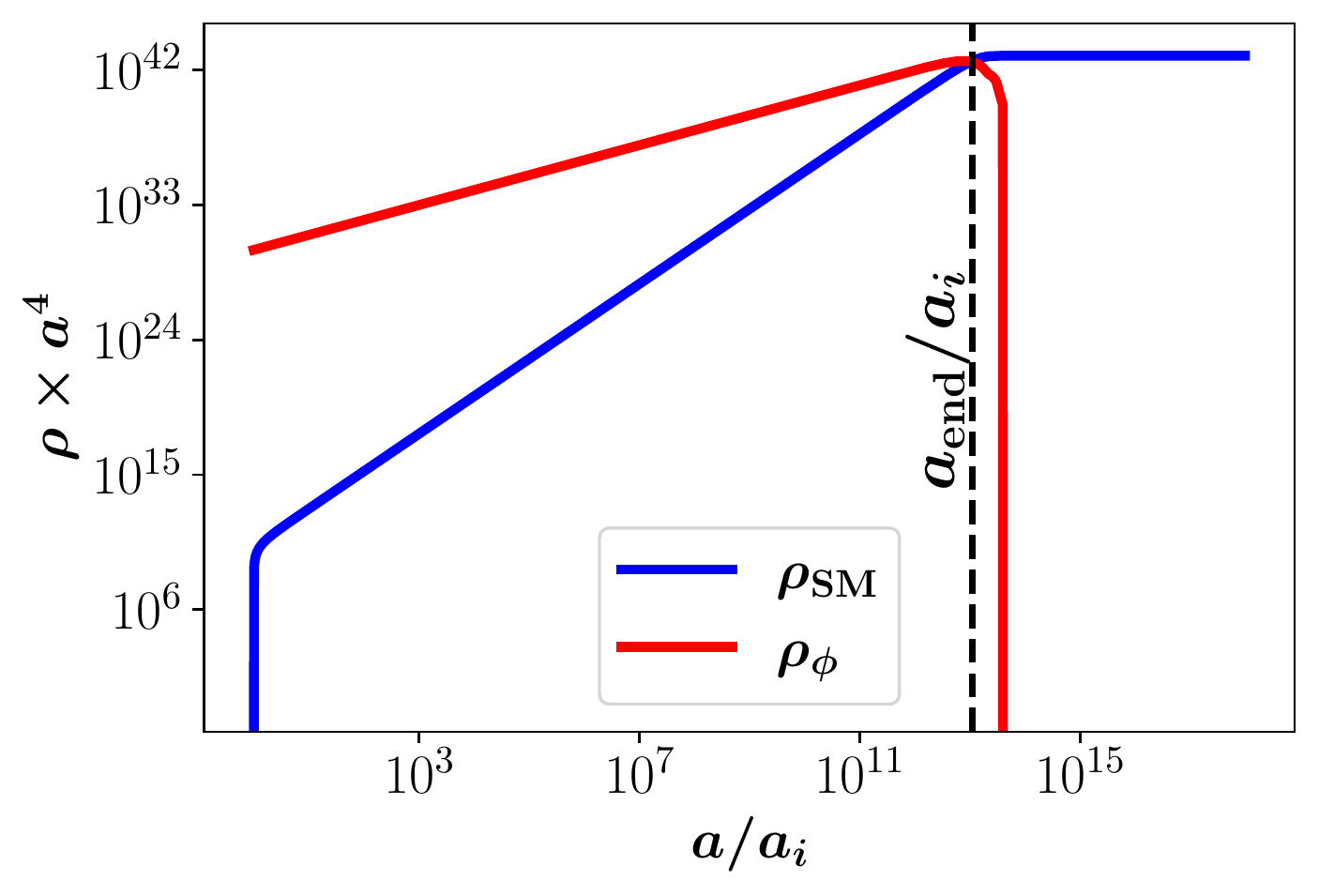}
         }
    \end{subfigure}
    \begin{subfigure}[\label{fig:Tinflaton}]{
         \centering
         \includegraphics[width=0.46\linewidth]{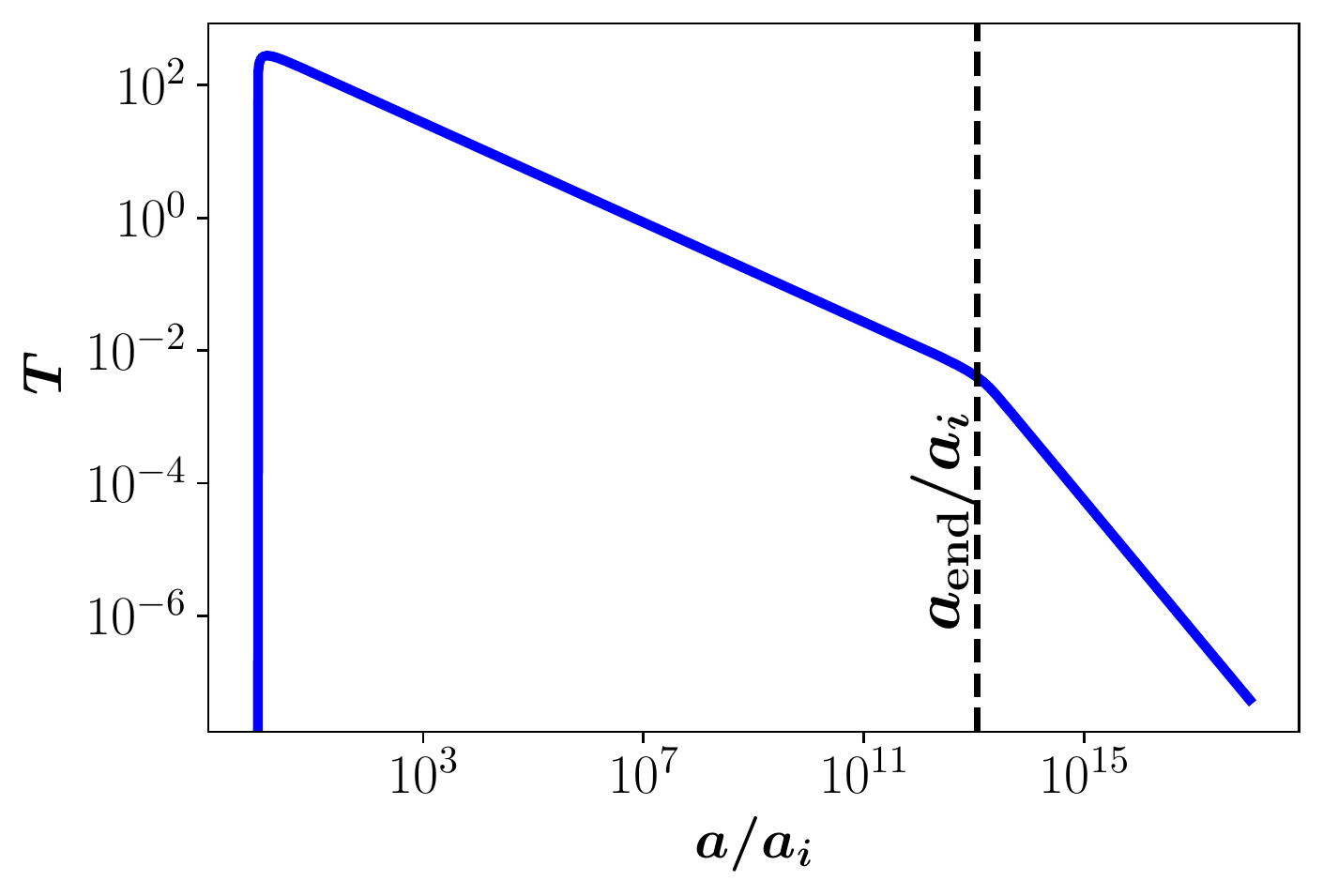}
         }
    \end{subfigure}
    \begin{subfigure}[\label{fig:Tinflaton2}]{
         \centering
         \includegraphics[width=0.46\linewidth]{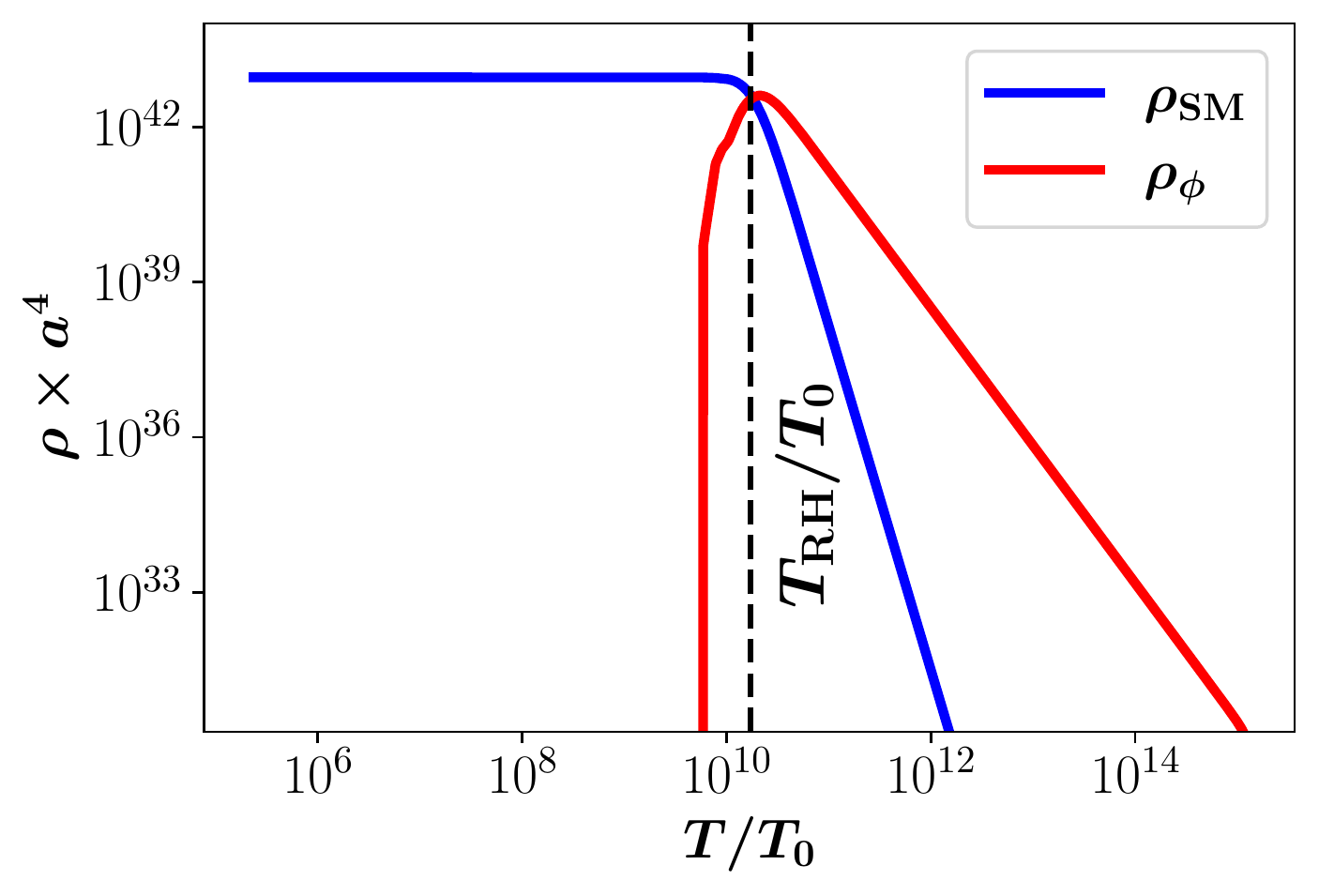}
         }
    \end{subfigure}
    \caption{Panel (a) shows the  $\phi$ field acting like an inflaton. The SM sector starts to grow with the evolution of $\phi$ until  $T_{RH}=4\times10^{-3}$ GeV is reached and $\phi$ decays.  In panel (b)  the temperature as a function of $a/a_i$ is shown.  The change of slope at $T_{RH}$ is  due to the decay of the  $\phi$ field. Energy density as function of $T$ is shown in (c), with $T_0=2.33\times 10^{-13}$ GeV.}
    \label{fig:phiinflaton}
\end{figure}

The $\phi$ field has interesting features. It acts like an inflaton when the initial energy density $\rho_\text{SM}$ is zero and then, generates a new epoch of reheating due to the decay term which transfers energy to the SM content until $T_{RH}$ is reached \cite{Giudice:2001,Maldonado:2019}.
At this temperature the  $\phi$ field decays completely. This effect is shown in Figure \ref{fig:phiinflaton}. 

For a less restrictive scenario, one assumes  a  non zero ratio between the energy density of 
$\phi$  and the energy density of the SM, at some initial scale factor $a_i$. That is, a non zero
value for the quantity
$$
\delta=\frac{\rho_\phi}{\rho_\text{SM}}\bigg{|}_{a_i}.
$$
In this case, this new field does not act like an inflaton anymore, 
but  we can observe a similar behavior growing up the energy density for the SM content meanwhile the $\phi$ field is decaying until $T_{\text{end}}$, 
which is the temperature when total  decay occurs \cite{Bernal:2019}.

\begin{figure}[h!]
    \centering
    \begin{subfigure}[  \label{fig:rhokappa}]{
         \centering
         \includegraphics[width=0.46\linewidth]{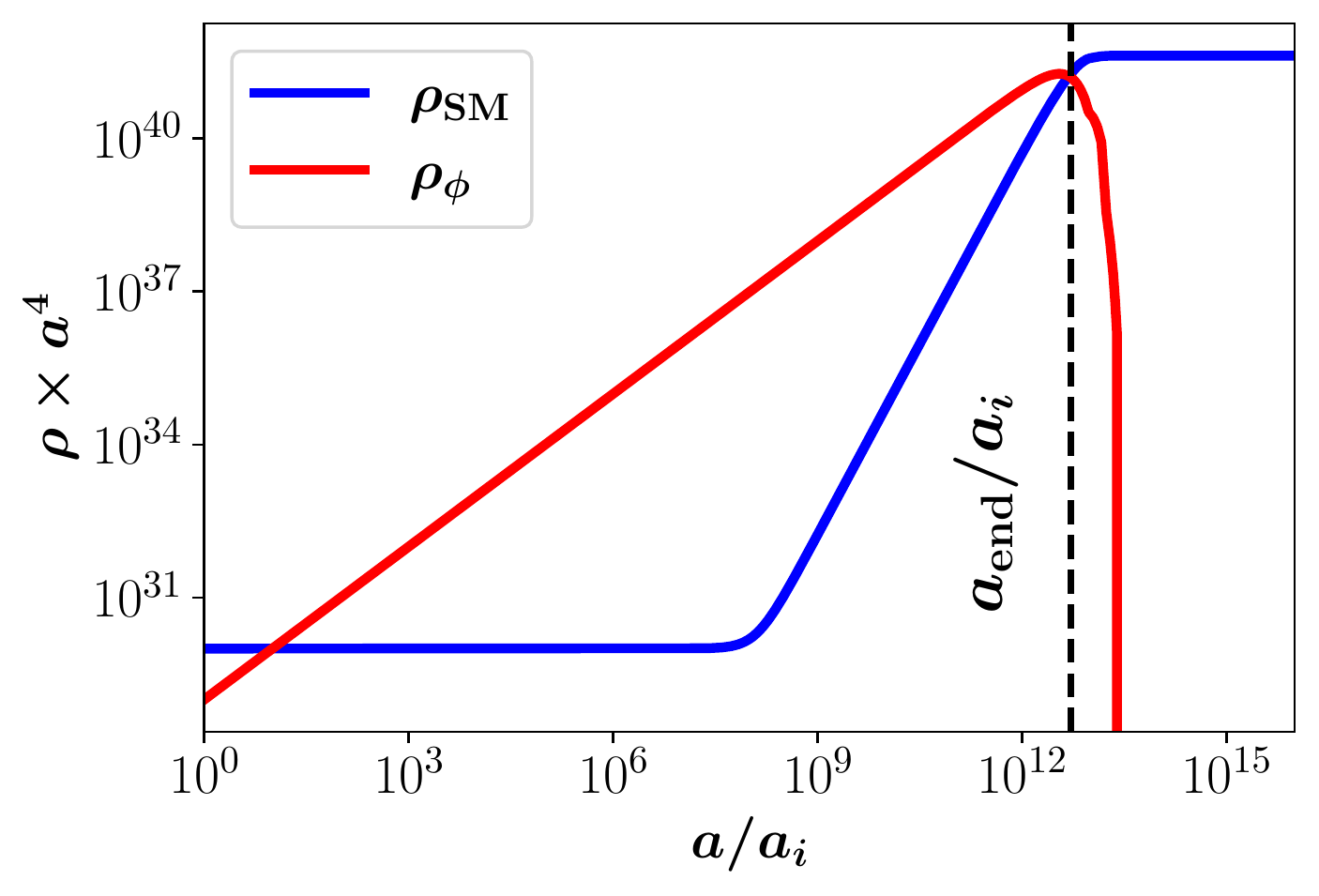}
    }
    \end{subfigure}
    \begin{subfigure}[ \label{fig:Tinkappa}]{
         \centering
         \includegraphics[width=0.46\linewidth]{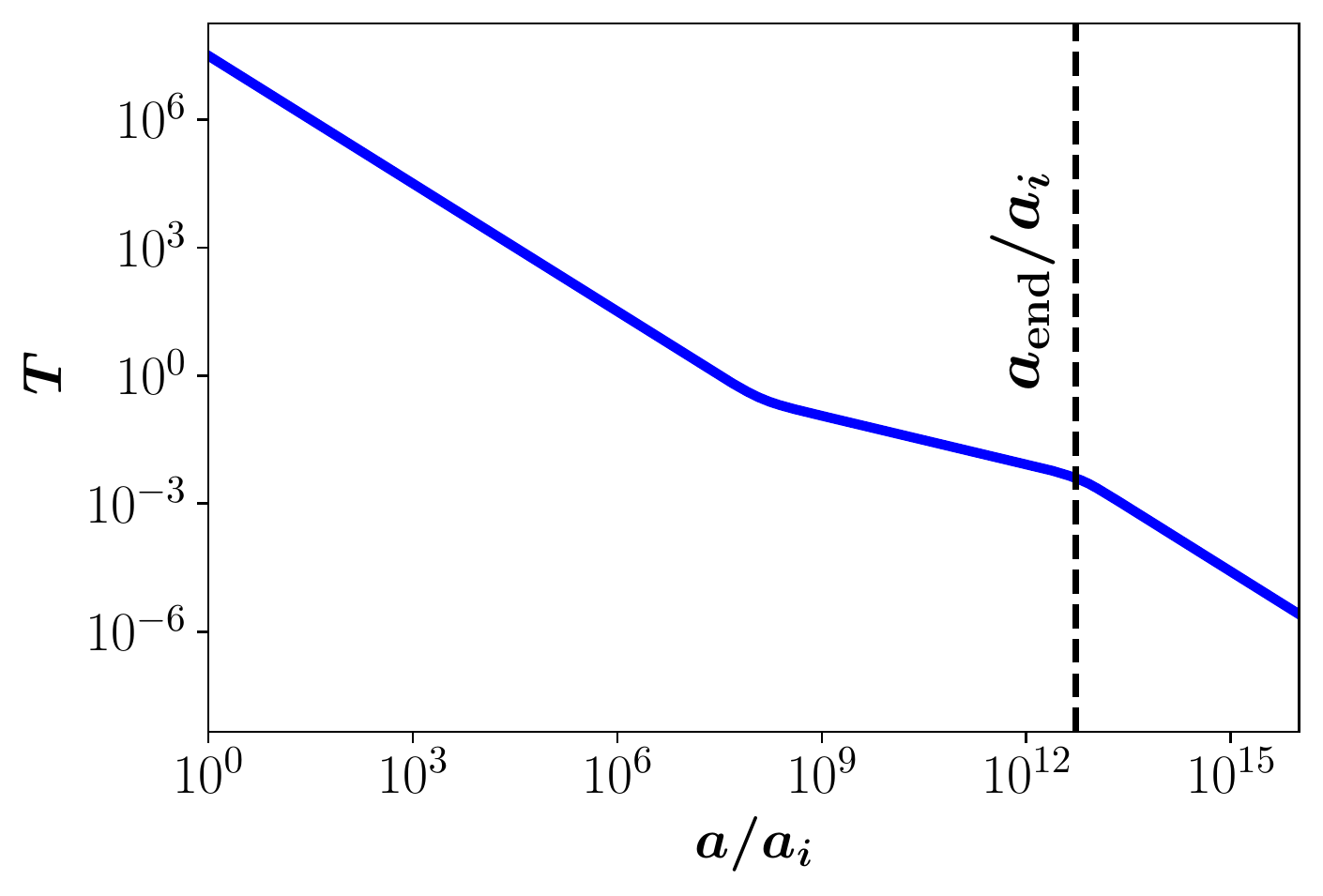}
     }
     \end{subfigure}
     \begin{subfigure}[ \label{fig:rhoTka}]{
         \centering
         \includegraphics[width=0.46\linewidth]{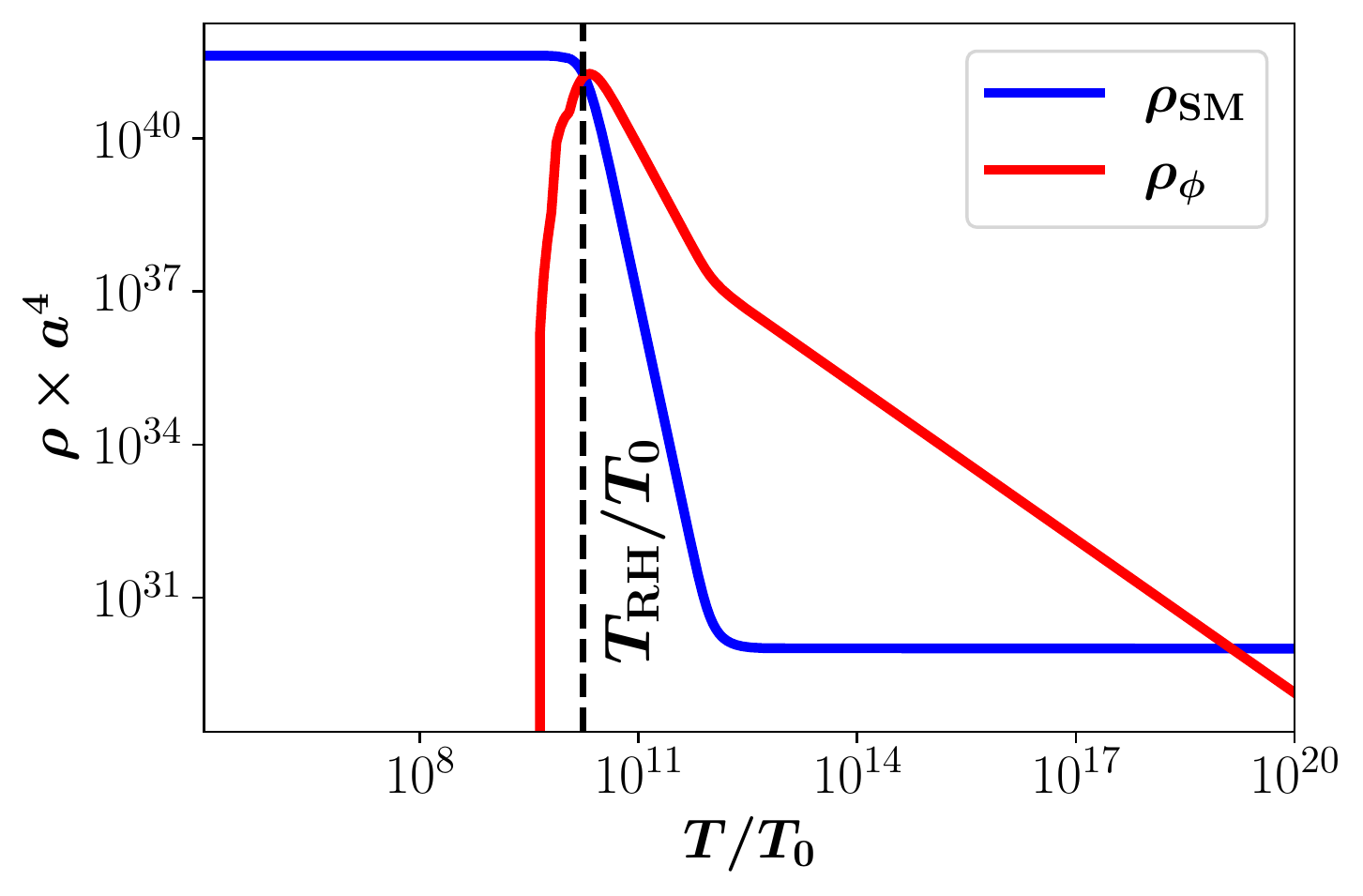}
     }
     \end{subfigure}
     \caption{Panel (a) shows the evolution of energy density of  $\phi$ field and the SM bath for  $\delta=10^{-1}$. $\Gamma$ is chosen so that $\phi$ decays   at $T_{RH}=4\times10^{-3}$ GeV. Panel (b) exhibits the temperature as a function of $a/a_i$. The slope changes  due to the effect the $\phi$ field which is decaying. The panel (c) show the relationship between the energy densities and the temperature, with $T_0=2.33\times 10^{-13}$ GeV the current temperature.}
     \label{fig:phikappa}
\end{figure}

It is interesting to note that the two-metric model 
in absence of matter also shows an inflaton-like behavior 
\cite{Falomir1}, but there the interaction provided by the deformation of the Poisson bracket 
structure is the responsible for such effect. We will show
that for the matter case, it is possible to reproduce also 
the behavior shown in Figure \ref{fig:phikappa}.

\section{The matter content}

We are interested in the study of energy density  evolution 
under the hypothesis that the evolution of matter in one patch 
is independent from the other and  particularized to  
barotropic perfect fluids characterized by the
presure  $p$ and the energy density $\rho$. 

Note that the LHS of  (\ref{eom2a})  and (\ref{eom2b})
are proportional to the spatial component of the Einstein tensor. 
Indeed, for a FLRW metric with scale
factor $a$ and spatial curvature $k_a$ (in the gauge $N=1$) the Einstein 
tensor reads
\begin{equation}
G_{ij} = -g_{ij}\left(2a\ddot{a}+\dot{a}^2+k_a\right),\quad G_{00} = \frac{3}{a^2}\left(\dot{a}^2+\kappa_a\right),
\end{equation}
and then, under the hypothesis previously explained, we propose the
following modification of the equations of motion in order to include
matter effects
\begin{eqnarray}
\label{matta}
2a\ddot{a}+\dot{a}^2 &=&\Lambda_a\,a^2-k_a+2\kappa\dot{b} - a^2 {p}_a,
\\
\label{mattb}
2b\ddot{b}+\dot{b}^2 &=&\Lambda_b\,b^2-k_b-2\kappa\dot{a} -  b^2{ p_b},
\\
\label{constmatt}
a\dot{a}^2+b\dot{b}^2&=&\frac{\Lambda_a}{3}\,a^3-k_a\,a+\frac{a^3}{3}{\rho}_a +
\nonumber
\\
& &\frac{\Lambda_b}{3}\,b^3 - k_b\,b+\frac{b^3}{3}{\rho}_b,
\end{eqnarray}
where  $p$ and $\rho$ are the pressure and energy density of the fluid 
(in $M_{\mbox{\tiny{Pl}} } = (8\pi G)^{-1/2}$ units), respectively, and 
index $a,b$ denotes the patch where they are defined.

A comment is in order here.  While the pressure terms in  (\ref{matta}) and 
(\ref{mattb}) trivially satisfy the hypothesis of 
local matter content, modifications of the constraint equation -- the term  
$a^3\rho_a+b^3\rho_b$ in (\ref{constmatt}) -- have not a unique form.
The most general  term modifying (\ref{constmatt}) must be a function $\rho^{(ab)}$ which 
satisfies  also the separability condition  $\rho^{(ab)} =\rho^{(a)} +\rho^{(b)}$,
since the   constraint in the present model turns out to be  the addition of the usual ones on each patch.
Indeed, for the FLRW metric with scale factor $a$, constraint reads  
${\cal C}_a = a\dot{a}^2 -  \frac{\Lambda_a}{3}\,a^3 +k_a\,a -\frac{a^3}{3}{\rho}_a =0 $, 
while for the present  two-metric model, the constraint reads ${\cal C}_a +{\cal C}_b =0$.

Previous characteristic is a consequence of  the fact that there is only 
one time for both patches (and then, only  one lapse function $N$) ensuring the time
reparametrization invariance.  Then, our choice of energy density term respects
the separability condition and it reproduces also the standard cosmological 
scenario if both patches are not connected, that  is $\kappa =0$.

The conservation law is obtained by taking the time derivative of 
the constraint  and  replacing the second derivatives of the scale factors
from (\ref{matta}) and (\ref{mattb}). For a general energy-density term 
$\rho^{(ab)}$ the continuity equation reads\footnote{This is a  notation abuse since $\rho^{(ab)}$ has not the dimensions of energy density} 
\begin{equation}
\label{conti}
\dot{\rho}^{(ab)} +\dot{a}\,a^2\,p_a+\dot{b}\,b^2\,p_b =0.
\end{equation}
Once we specify the function $\rho^{(ab)}$ to our choice in (\ref{constmatt}), previous equation turn 
out to be
\begin{eqnarray}
\label{contifinal}
a^3\left[\dot{\rho}_a+3H_a^2\left(\rho_a+p_a\right)\right]+
b^3\left[\dot{\rho}_b+3H_b^2\left(\rho_b+p_b\right)\right]&=&0,
\nonumber
\\
&&
\end{eqnarray}
with $H_a=\dot{a}/a$ and $H_b =\dot{b}/b$, the Hubble parameters on each patch.

To summarize, in the present approach where matter on patches $a$ and $b$ are characterized by their
pressure and energy density (on each patch), the evolution of the scale factors are given by equations 
(\ref{matta}) to (\ref{constmatt}) from which the conservation equation (\ref{contifinal}) follows.

\section{Barotropic mater in the early universe}

We will analyze the effects of the matter presence for the case in which 
fluids on $a$ and $b$  satisfy the barotropic condition
\begin{equation}
\label{baro}
p_a=\omega_a\,\rho_a,\quad p_b = \omega_b\,\rho_b.
\end{equation}
In the forthcoming analysis,  the  contributions from cosmological constant
will be neglected since we are interested in the early stage of the evolution
of the universe, which is an interesting scenario for different physical phenomena like Dark Matter production \cite{Giudice:2001,Maldonado:2019,Bernal:2019,Arias:2019} or gravitational waves \cite{Bernal2}, among others. Also, we set $k_a=0=k_b$, the favored scenario consistent
with cosmological data \cite{Zyla:2020zbs}.  Note also that, in spite of  the
choice $\Lambda_a=0=\Lambda_b$, a sort of cosmological constant term is always
present due to the effects of a non zero value of $\kappa$  \cite{Falomir1}.
The equations of evolution, with previous choices, turn out to be
\begin{eqnarray}
\label{baroa}
2\frac{\ddot{a}}{a}+H_a^2 +\omega_a\,\rho_a  &=& ~ 2\kappa\,\frac{\dot{b}}{a^2},
\\
\label{barob}
2\frac{\ddot{b}}{b}+H_b^2 +\omega_b\,\rho_b &=& - 2\kappa\,\frac{\dot{a}}{b^2},
\\
\label{baroconst}
a^3\left[ H_a^2 -\frac{1}{3}\rho_a\right] &+&
b^3\left[ H_b^2 -\frac{1}{3}\rho_b\right] 
= 0,
\\
 && \nonumber
\end{eqnarray}
while the continuity equation read
\begin{eqnarray}
\label{continuityfinal}
a^3& &\left[\dot{\rho}_a+3\rho_a H_a^2\left(\omega_a+1\right)\right]+
\nonumber
\\
b^3&&\left[\dot{\rho}_b + 3\rho_b H_b^2\left(\omega_b+1\right)\right] = 0.
\end{eqnarray}

In the present model, we will look for solutions of (\ref{baroconst}) respecting 
the separability hypothesis and then, we look for solutions which are also the solutions of 
\begin{eqnarray}
\label{consta}
   H_a^2 -\frac{\rho_a}{3}&=&0,
    \\
    \label{constb}
    H_b^2 -\frac{\rho_b}{3}&=&0.
\end{eqnarray}
The  time derivative of previous equations give rise to the following conditions
\begin{eqnarray}
\label{continuitya}
    a^3 \left(\dot{\rho}_a+3H_a(\omega_a+1)\rho_a\right) &=&  6\kappa \,\dot{a}\dot{b},
    \\
    \label{continuityb}
    b^3 \left(\dot{\rho}_b+3H_b(\omega_b+1)\rho_b\right) &=& - 6\kappa \,\dot{a}\dot{b},
\end{eqnarray}
which, when added, turn out to be  (\ref{continuityfinal}).

Comparing  (\ref{continuitya}) and (\ref{continuityb}) with (\ref{radNSC}) and (\ref{phiNSC})
for the case of NSC, we observe  the   similar source-sink behavior due to the $\kappa$ 
term in the two-metric model.  However, the decaying constant $\Gamma$ is now time dependent.
Moreover,  one can rewrite (\ref{continuitya})  and (\ref{continuityb}) as
\begin{eqnarray}
\label{identifhi}
     \dot{\rho}_a+3H_a(\omega_a+1)\rho_a &=&  \Gamma_a\, \rho_b,
    \\
\dot{\rho}_b + 3H_b(\omega_b+1)\rho_b &=& -\Gamma_b\,\rho_b,
\end{eqnarray}
with 
\begin{equation}
    \label{Gammaseq}
    \Gamma_a =2\kappa\frac{b}{a^2}\,\delta^{-1/2},\quad \Gamma_b =2\kappa\frac{a}{b^2}\,{\delta^{-1/2}},
\end{equation}
and $\delta = \rho_b/\rho_a$. The decay functions $\Gamma$ satisfy $a^3\Gamma_a-b^3\Gamma_b =0$.
In this sense, the two-metric model with matter can be understood as an extension of NSC.

In the following sections we will study  the numerical solutions of the set
of equations (\ref{consta}) to (\ref{continuityb}) in two cases. For both scenario 
the energy density in $a$ patch will have a radiation-like barotropic equation so that we can 
compare with with NSC, (we will refer this content as relativistic)  while the patch $b$ contains relativistic matter  in one case and 
non-relativistic, in the other. Even though the 
numerical solutions found are functions of time, it is convenient to 
express results in terms of temperatures.

The temperature dependence is incorporated by noticing that in patch $a$, where
relativistic matter dominates, the following relation holds
\begin{equation}
    \rho_a =\frac{\pi^2}{30}g_* T^4,
\end{equation}
with $g_*$ the number of massless degrees of freedom.

\subsection{Patch $b$ filled with non-relativistic matter}
In this case, as we previously discussed, patch $a$ is filled with relativistic
matter while the energy content of $b$ is non-relativistic. Then $\omega_a=1/3$ and $\omega_b=0$
and the set of equations (\ref{consta}) to  (\ref{continuityb}) to determine time evolution of 
scale factors and  energy densities  are 
\begin{eqnarray}
\label{radmatttime}
\dot{a}&=&\sqrt{\frac{\rho_a}{3M_p^2}}\,a,
\nonumber
\\
\dot{b}&=&\sqrt{\frac{\rho_b}{3M_p^2}}\,b,
\nonumber
\\
\dot{\rho}_a&=&-\frac{4}{\sqrt{3M_p^2}}(\rho_a)^{3/2}
+ 2 \kappa M_p\sqrt{\rho_a \rho_b}\,\frac{b}{a^2},
\nonumber
\\
\dot{\rho}_b&=&-\frac{1}{\sqrt{3M_p^2}}(\rho_b)^{3/2}
- 2 \kappa M_p\sqrt{\rho_a \rho_b}\,\frac{a}{b^2},
\end{eqnarray}
where we have restored the Planck mass constant.

\begin{figure}[h!]
    \centering
    \subfigure[\label{fig:a}]{
         \includegraphics[width=0.46\linewidth]{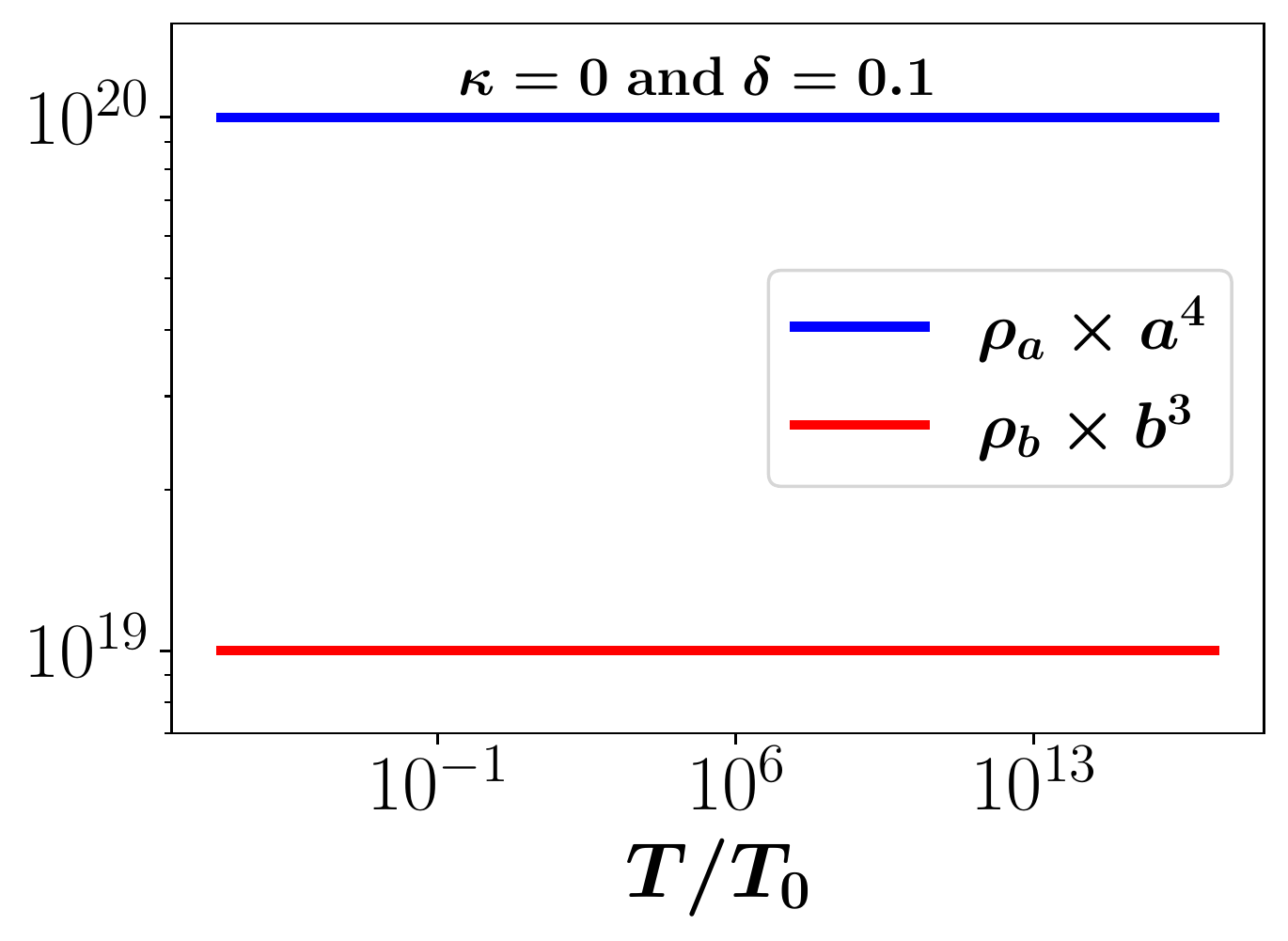}
     }
     \subfigure[\label{fig:b}]{
         \centering
         \includegraphics[width=0.46\linewidth]{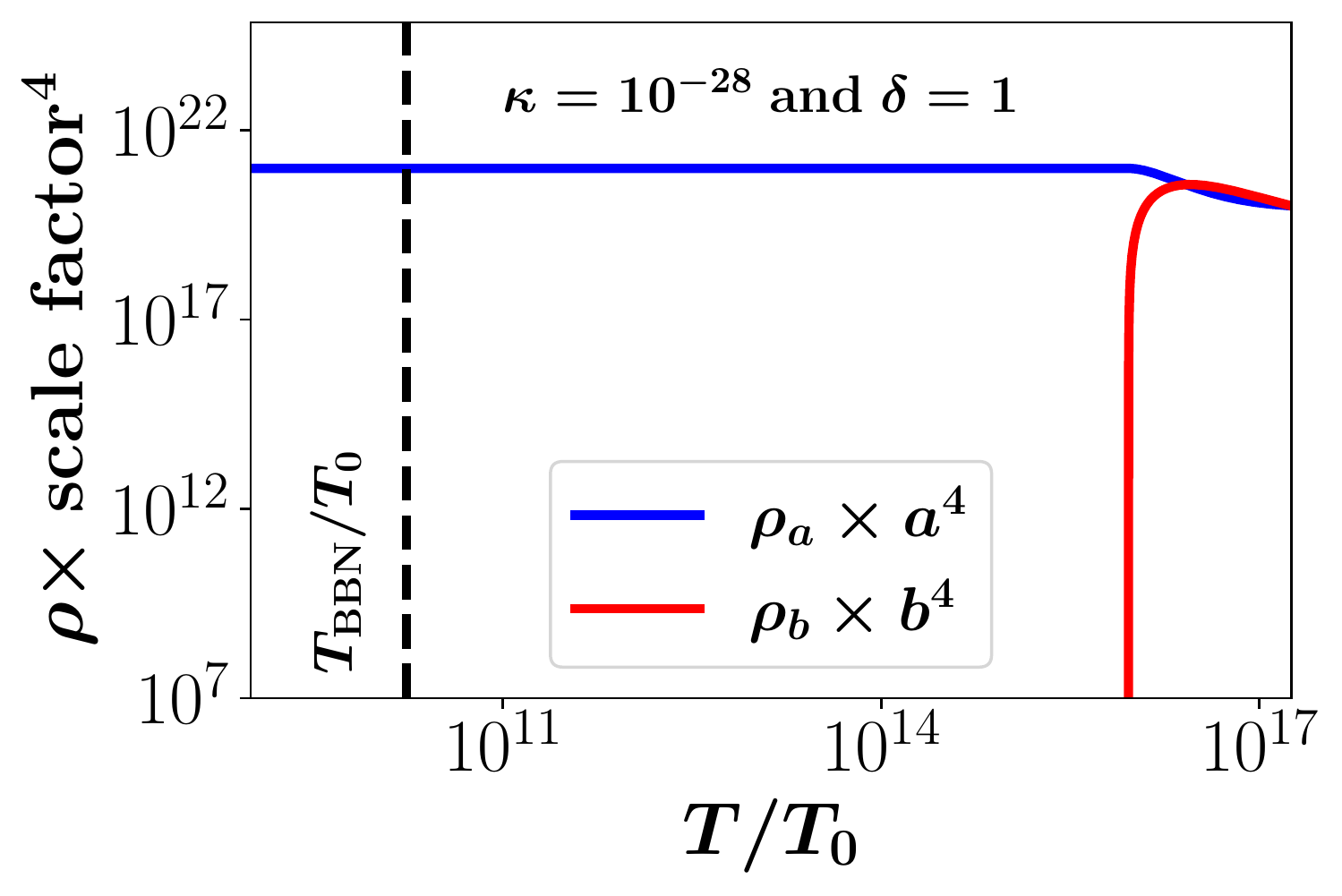}
     }
      \subfigure[\label{fig:c}]{
         \includegraphics[width=0.46\linewidth]{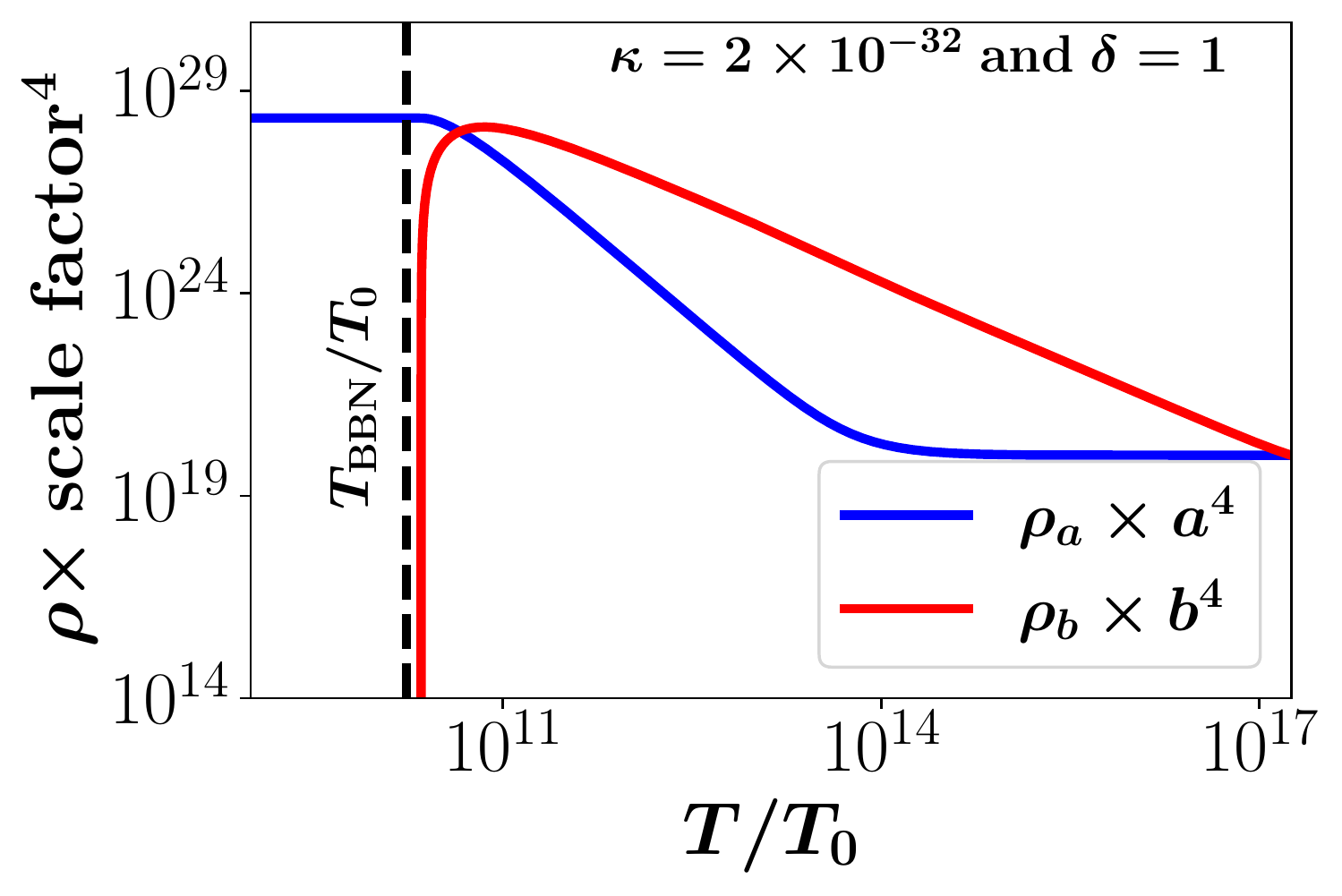}
     }
     \subfigure[\label{d}]{
         \centering
         \includegraphics[width=0.46\linewidth]{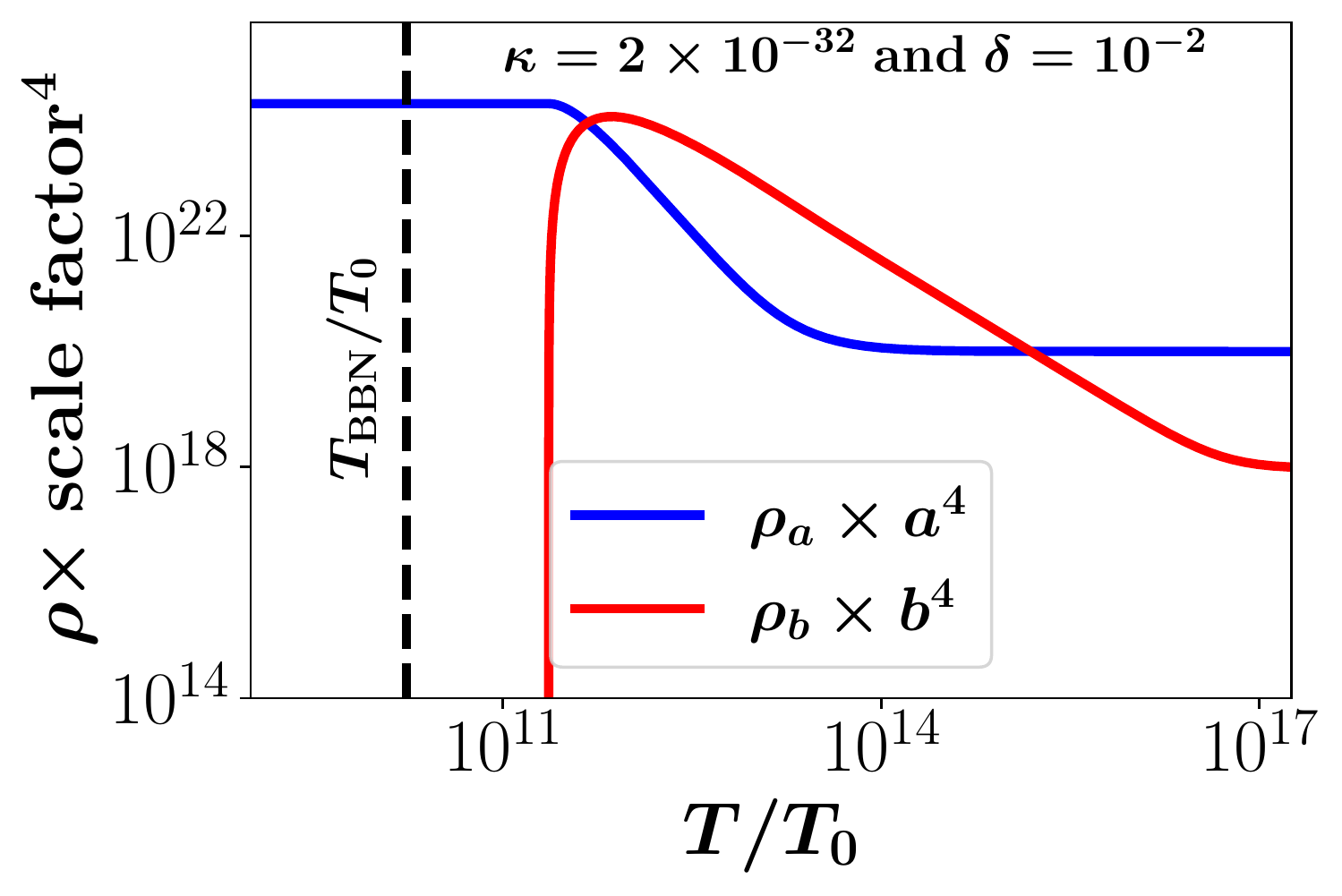}
     }
      \subfigure[\label{fig:e}]{
         \includegraphics[width=0.46\linewidth]{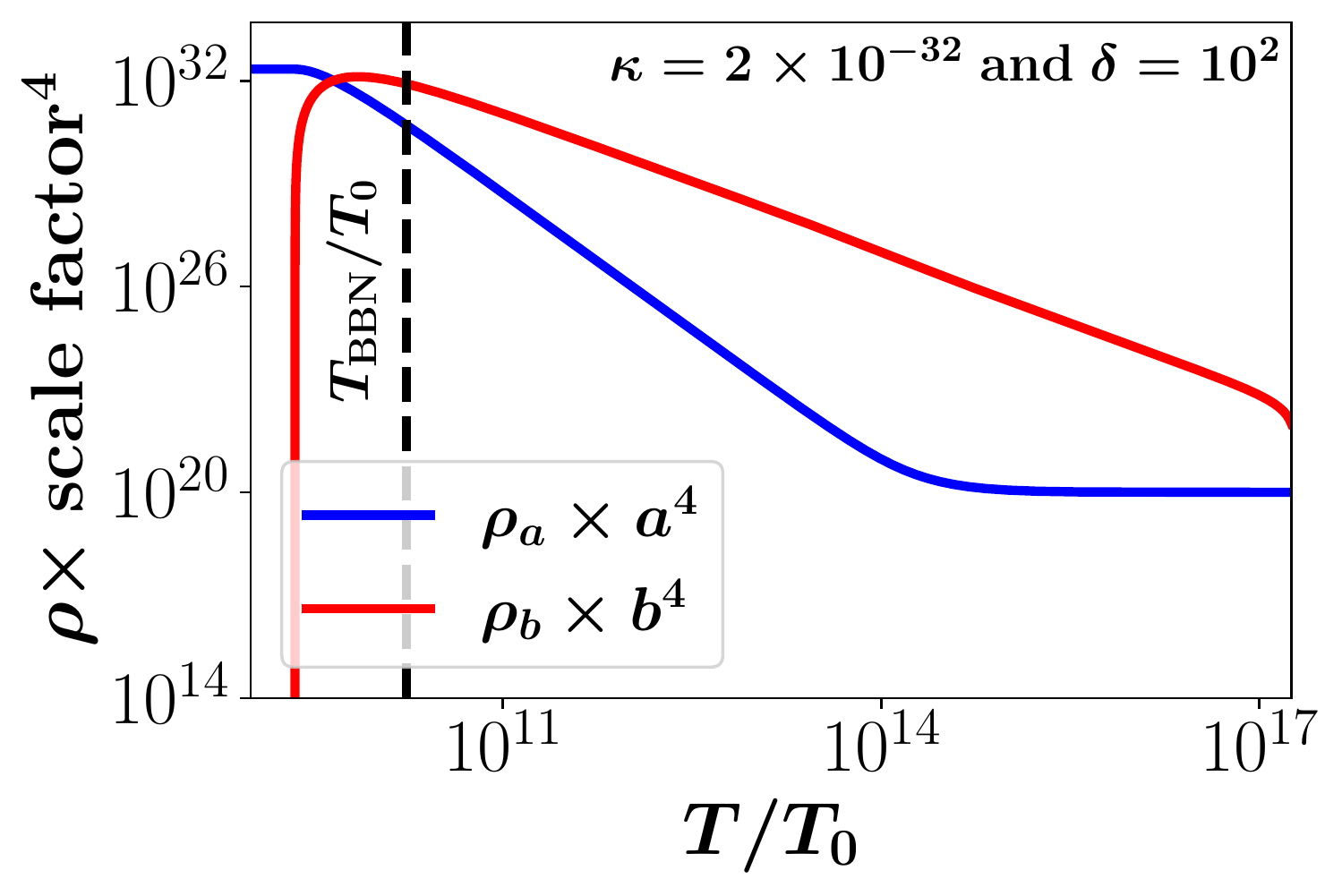}
     }
     \caption{Evolution of energy density for relativistic matter in $a$ and non-relativistic matter in $b$ as function of the temperature $T/T_0$. $T_0$ is the CMB temperature at present ($T_0=2.33\times 10^{-13}$ GeV). The dashed line indicates    the Big Bang Nucleosynthesis temperature at which $\rho_b$ should vanish. For all cases the initial density  $\rho_a=10^{20}$ GeV. Panel  (a) shows the case $\kappa =0$. Panels (b) and (c) are the solutions for $\delta=1$ and different values of $\kappa$.      The panels (d) and (e) show the evolution for $\delta\neq 1$ and same value of $\kappa$.}
     \label{fig:casematrad1e20}
\end{figure}

\begin{figure}[h!]
    \centering
     \subfigure[\label{fig:2}]{
         \centering
         \includegraphics[width=0.46\linewidth]{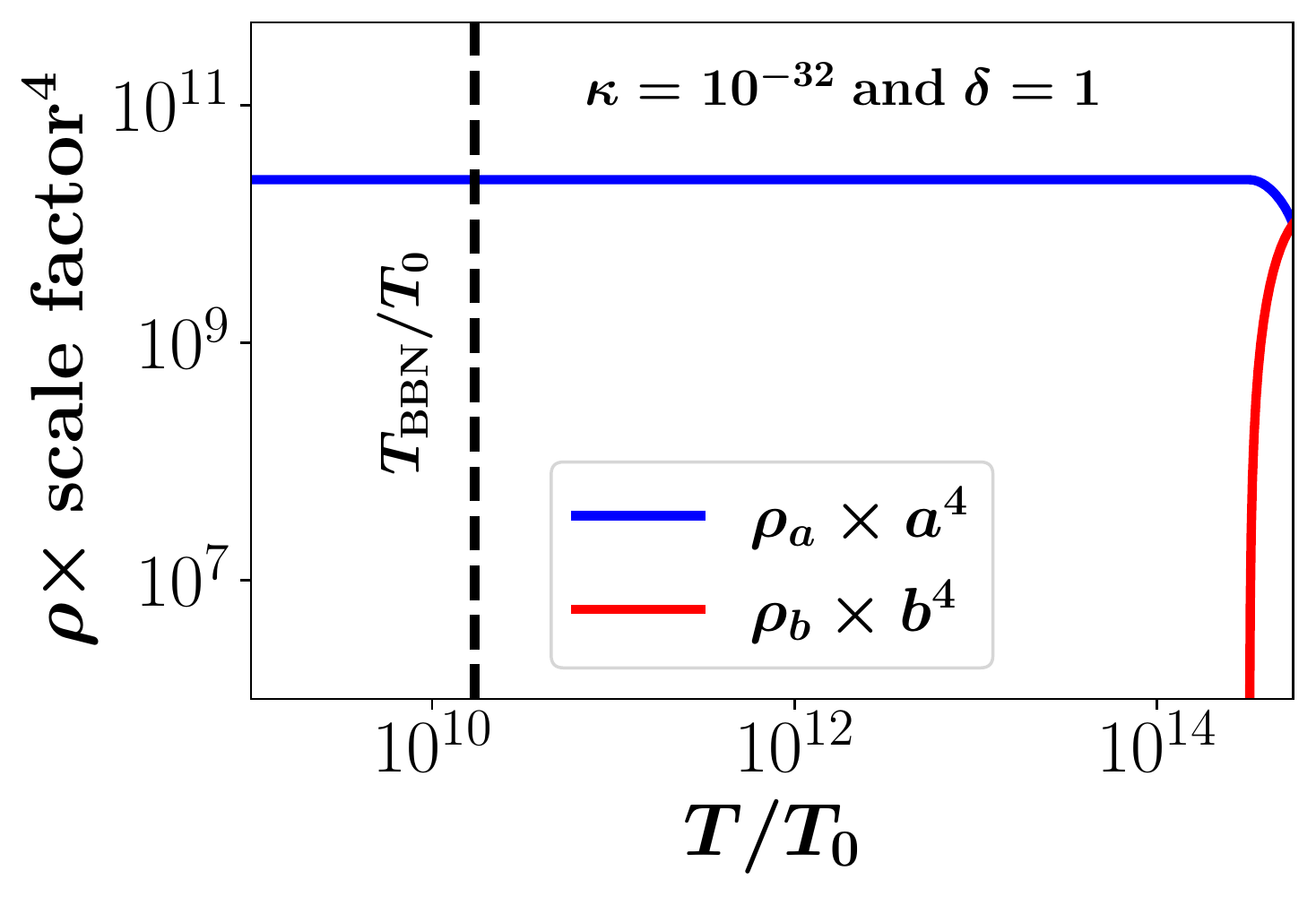}
     }
      \subfigure[\label{fig:4}]{
         \includegraphics[width=0.46\linewidth]{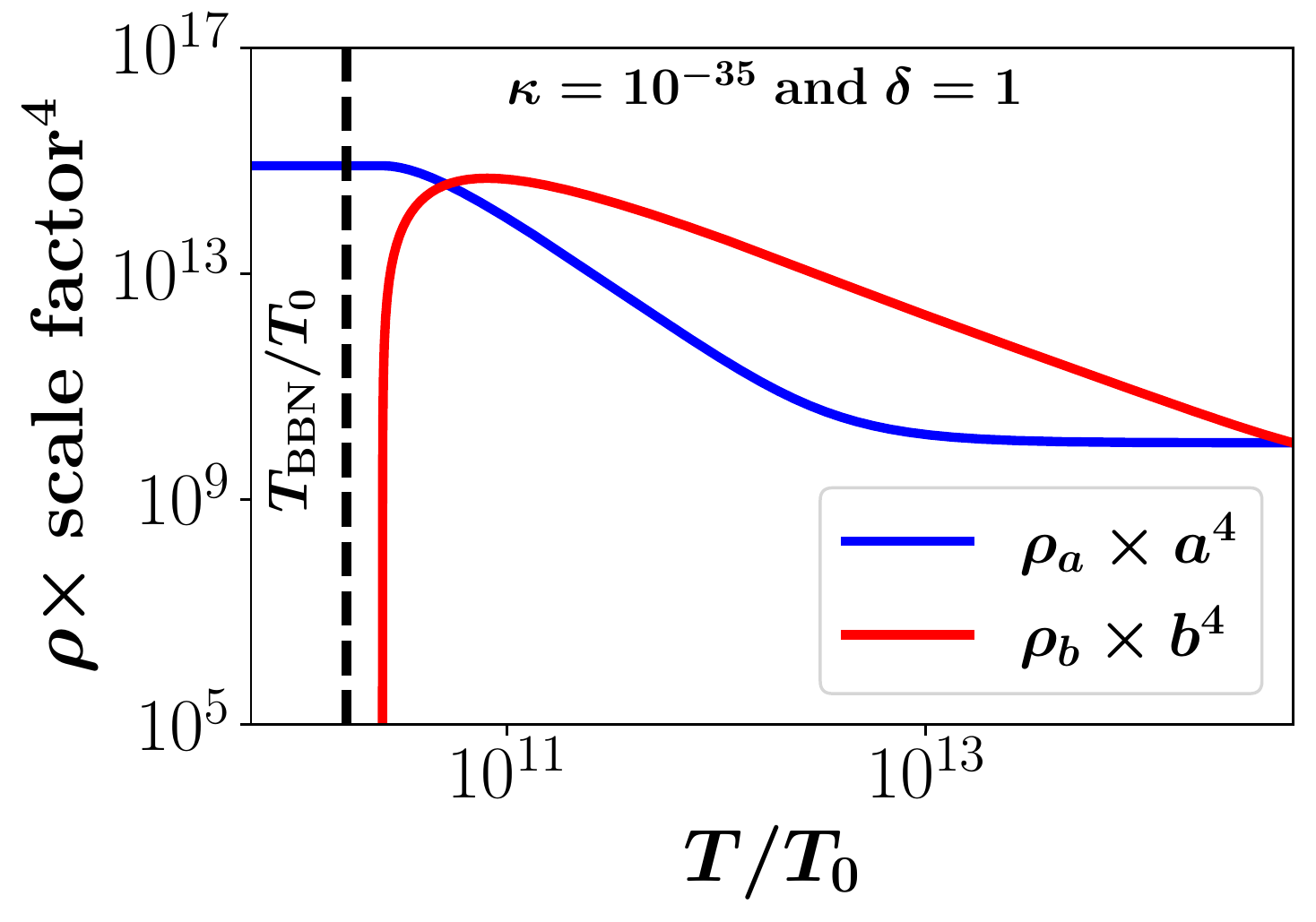}
     }
     \caption{Evolution of energy density for different values of $\kappa$ and initial condition  $\rho_a=10^{10}$ GeV with $\delta=1$.}
     \label{fig:casematrad1e10}
\end{figure}

\begin{figure}[h!]
    \centering
     \subfigure[\label{fig:6}]{
         \centering
         \includegraphics[width=0.46\linewidth]{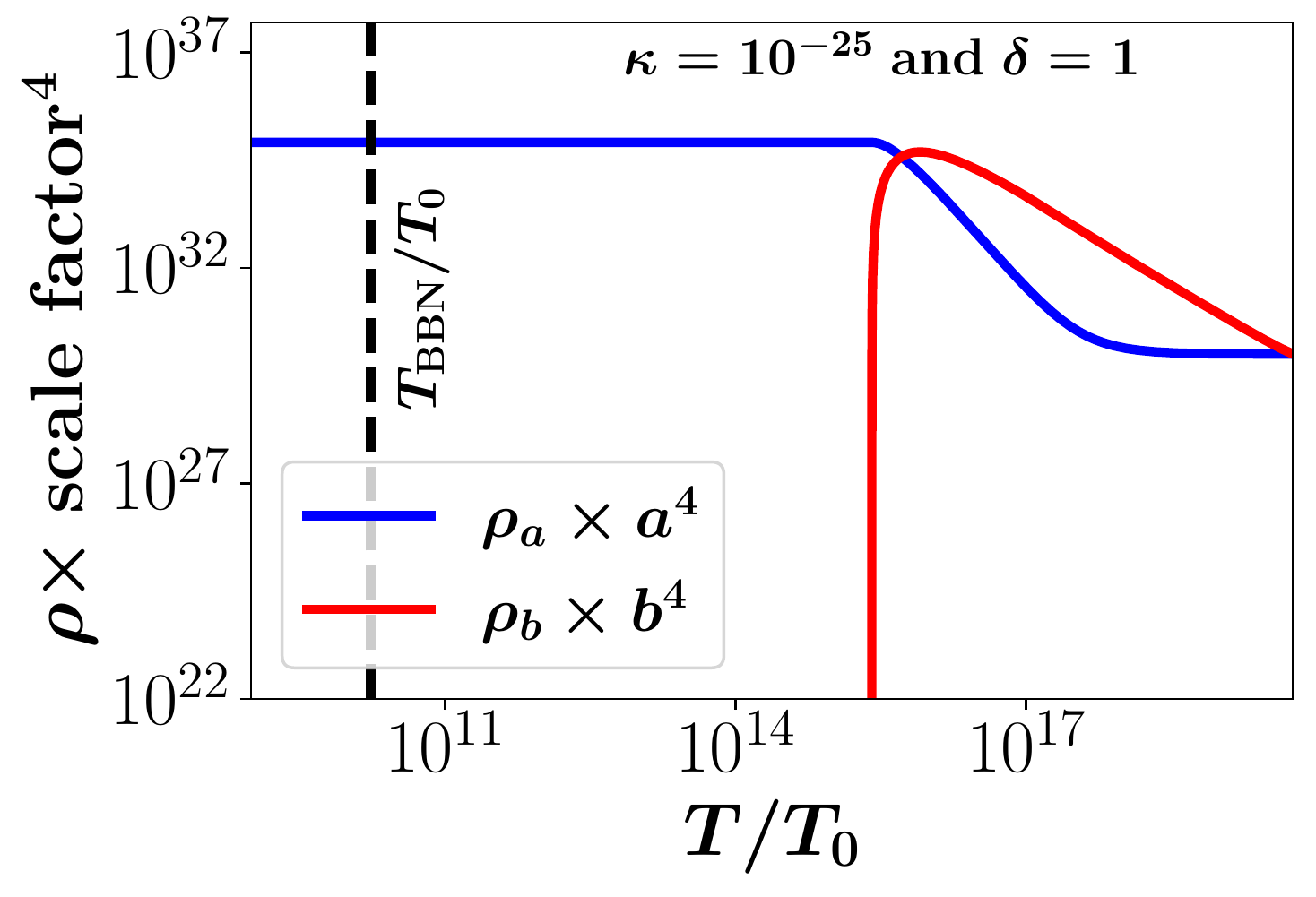}
     }
      \subfigure[\label{fig:8}]{
         \includegraphics[width=0.46\linewidth]{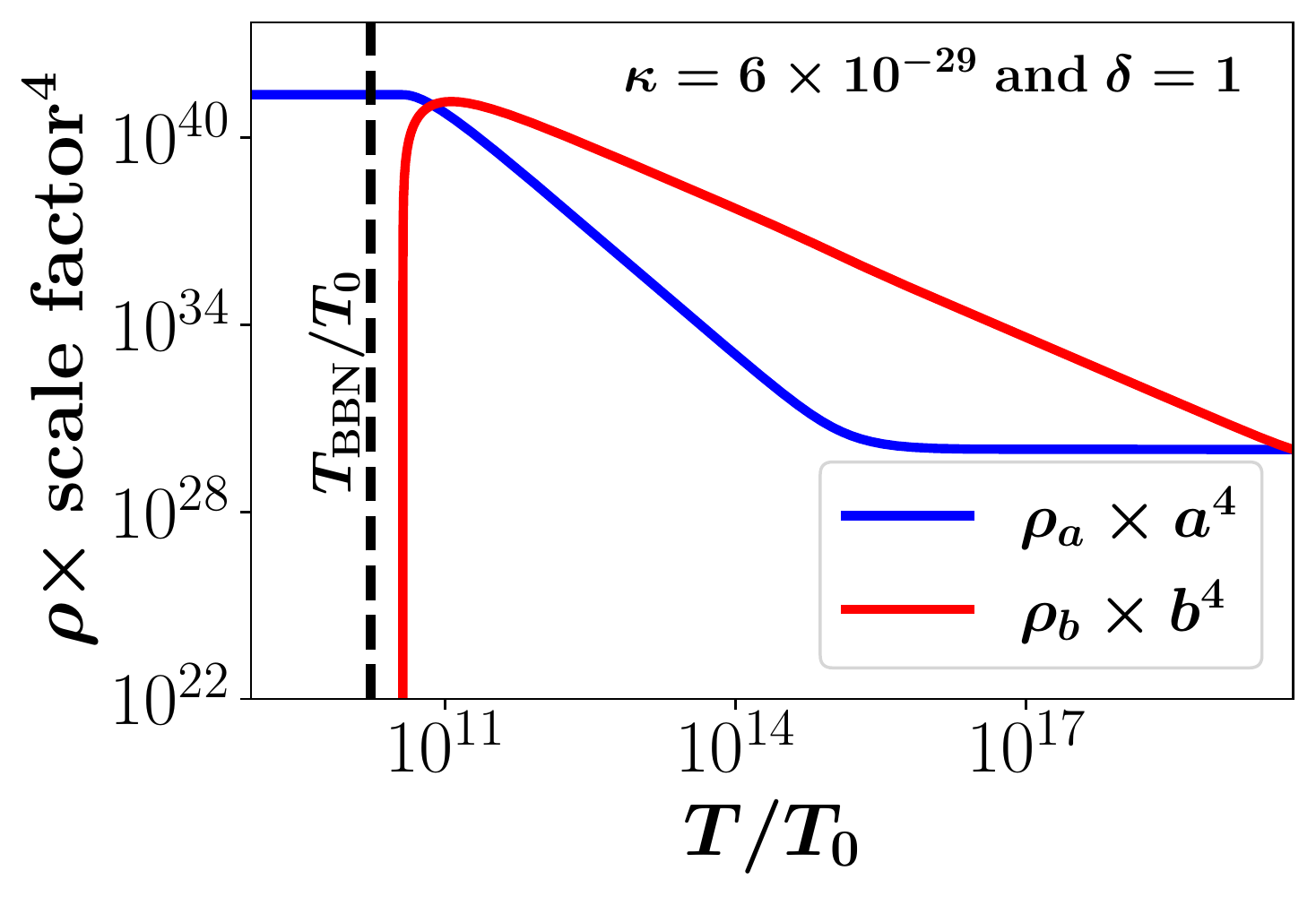}
     }
     \caption{Evolution of energy density for different values of $\kappa$ and initial condition  $\rho_a=10^{30}$ GeV with   $\delta=1$.}
     \label{fig:casematrad1e30}
\end{figure}

Numerical results for  the energy density evolution as function of temperature are
shown in Figures \ref{fig:casematrad1e20}, \ref{fig:casematrad1e10} and \ref{fig:casematrad1e30}.
The quantities of interest, as function of temperature, are $\rho\times$(scale factor)$^\ell$, for
some power $\ell$. 

In all cases with $\kappa\neq0$ we observe a {\it drain effect}, namely, the energy density of sector $b$
decrease until it vanishes,
while the energy density in $a$ increases. The temperature at which the total drain occurs 
depends on the value of $\kappa$ as well as the ratio $\delta$ at initial time. This is consistent
with the interpretation of source-sink system given by (\ref{identifhi}).

The dashed line marks the ratio $T_{BBN}/T_0$ at which  the drain of the energy content of $b$ should end. 
That is the drain must happen, at most, at the temperatures of the order of the temperature 
of Big Bang Nucleosynthesis ($T_{BBN}$) or higher than  $T_{BBN}$.

The panel (a) in Figure  \ref{fig:casematrad1e20} shows the situation for $\kappa=0$ in order to 
check that the systems are decoupled in such case and the energy densities evolve as it is expected
for radiation and non-relativistic matter. Thas is, $\rho_a\propto a^{-4}$ and $\rho_b\propto b^{-3}$.

Panels (b) and (c) of Figure  \ref{fig:casematrad1e20} show the case of initial ratio $\delta = 1$ and 
it is possible to observe that the decay of $b$ happens at higher temperatures (compared with $T_{BBN}$)
as $\kappa$ increases. Indeed,   it is enough to have $\kappa \gtrsim 10^{-32}$ in order to have a complete decay 
of energy content of $b$ sector at $T_{BBN}$. This is consistent with
the fact that $\Gamma_a$ in (\ref{identifhi}) is proportional to $\kappa$.

Let us take now the  value of $\kappa$ so that the total drain occurs at the desired temperature
for a symmetric initial density condition ($\delta =1$), situation shown in panel (c).
The effect of initial condition   $\delta >1$  and $\delta <1$ can be observed in panels (d) and (e)
in the same figure. We observe in panel (e) that when $b$ patch has more energy to drain (compared with the energy of 
$a$ patch) at the initial time, the complete process takes a longer time, so that the 
total decay of energy in patch $b$  happens at temperatures smaller than $T_{BBN}$, which is  an unfavorable scenario. 

In all previous cases the initial value of  energy density  in $a$ is $\rho_a = 10^{20}$ GeV. The effect
of a different initial condition for $\rho_a$  has been also addressed and the results are shown in  Figures \ref{fig:casematrad1e10} and
\ref{fig:casematrad1e30}. In the first, the initial value of energy density is  $\rho_a = 10^{10}$ GeV
while it is $\rho_a = 10^{30}$ GeV in the second.  For both cases we have chosen $\delta =1$. We conclude 
that the value of $\kappa$ for which the total drain happens at the desired temperature $T_{BBN}$ decreases
as the initial density $\rho_a$ decreases, what is consistent with our previous
result in Figure  \ref{fig:casematrad1e20}, panels (b) and (c).

To summarize, the temperature at which the energy density of sector $b$ vanishes, producing an
increment of the energy content in sector $a$ (a source-sink effect) depends on the values of 
$\kappa$, the initial value of the energy density in sector $a$\footnote{Naturally, it depends on the 
initial value of $\rho_b$ through $\delta$} and $\delta$. Large values of $\kappa$ produces a fast
decay of $\rho_b$ while large  values of initial $\rho_a$ slow down the decay rate. For a fixed 
$\kappa$, instead, large values of initial $\rho_b$ also slow down the decay rate. 

In the following section we will analyze the case in which the sector $b$ has relativistic matter
also and we will show that previous conclusions are also valid for such case.

\subsection{Patch $b$ filled with relativistic matter.}
In this case, the patches  $a$ and $b$ contain  relativistic
matter ($\omega_a=\omega_b=1/3$). The set of equations (\ref{consta}) to 
(\ref{continuityb}) to determine time evolution of scale factors and the energy 
density (with the Planck mass restored) are 
\begin{eqnarray}
\label{radradtime}
\dot{a}&=&\sqrt{\frac{\rho_a}{3M_p^2}}\,a,
\nonumber
\\
\dot{b}&=&\sqrt{\frac{\rho_b}{3M_p^2}}\,b,
\nonumber
\\
\dot{\rho}_a&=&-\frac{4}{\sqrt{3M_p^2}}(\rho_a)^{3/2}
+ 2 \kappa M_p\sqrt{\rho_a \rho_b}\,\frac{b}{a^2},
\nonumber
\\
\dot{\rho}_b&=&-\frac{4}{\sqrt{3M_p^2}}(\rho_b)^{3/2}
- 2 \kappa M_p\sqrt{\rho_a \rho_b}\,\frac{a}{b^2}.
\end{eqnarray}

\begin{figure}[h!]
    \centering
     \subfigure[\label{fig:rr2}]{
         \centering
         \includegraphics[width=0.46\linewidth]{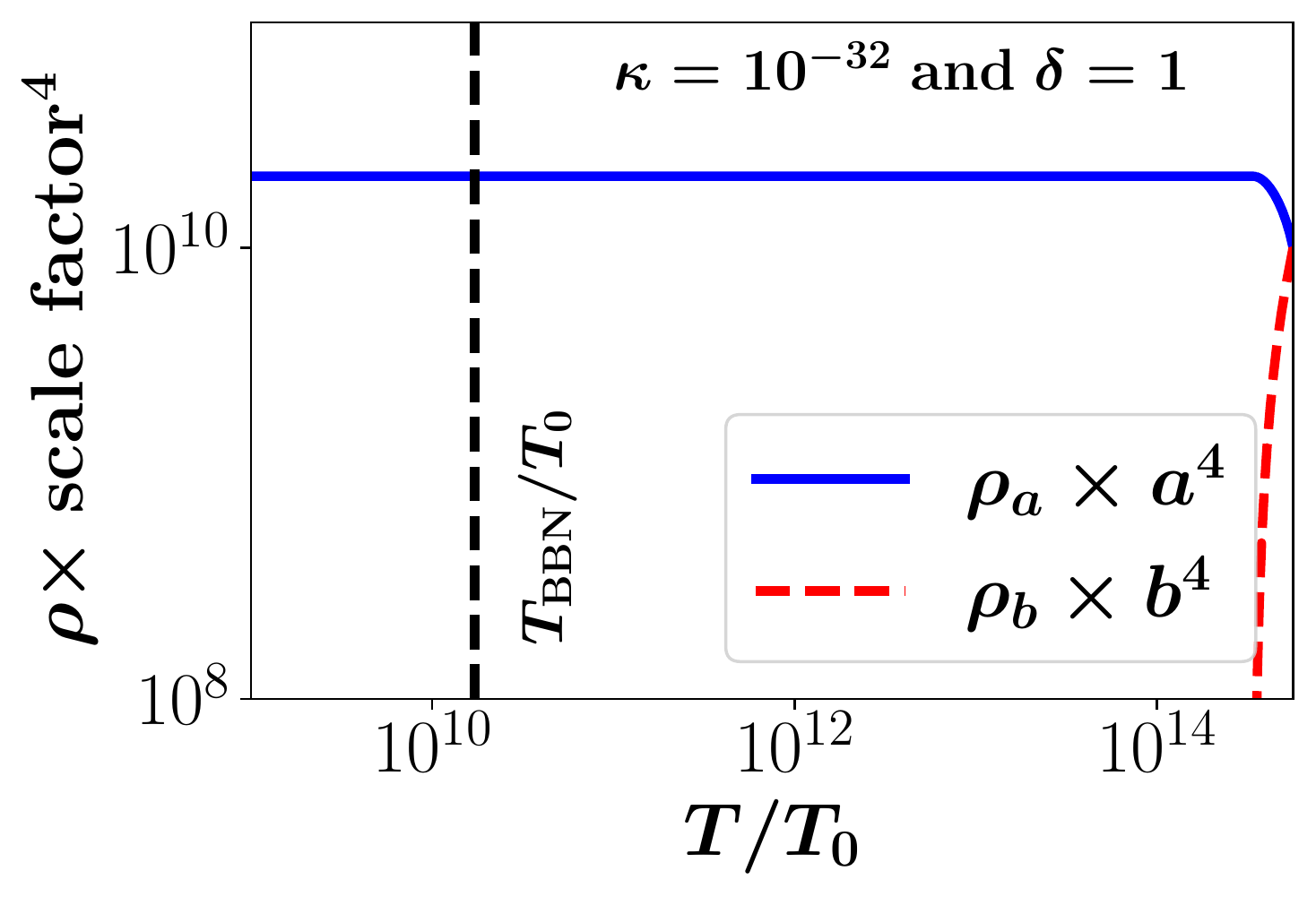}
     }
      \subfigure[\label{fig:rr4}]{
         \includegraphics[width=0.46\linewidth]{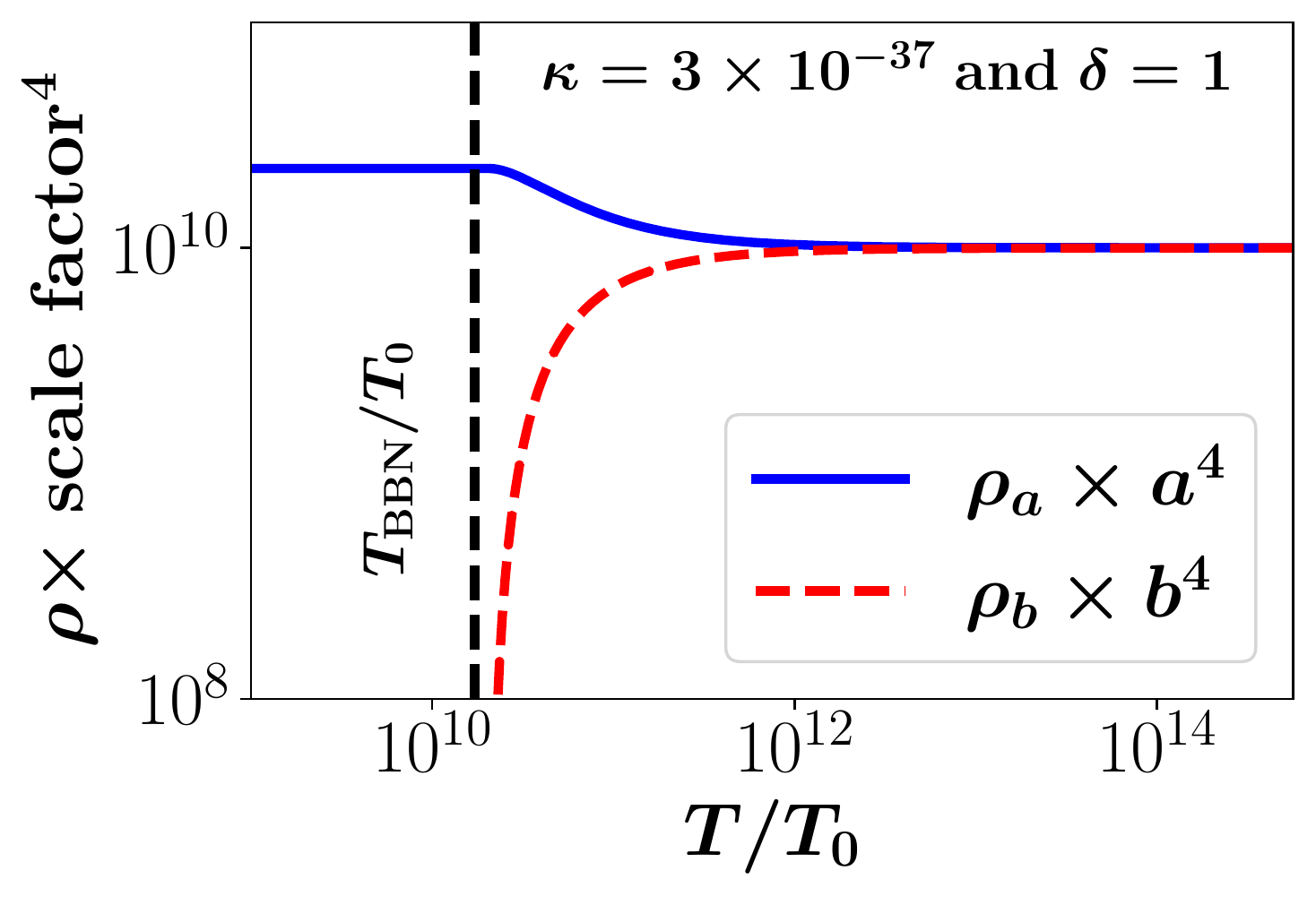}
     }
     \caption{Evolution of radiation content in patch  $a$ and $b$ as function of the temperature. Panels (a) and (b) show the energy densities
     evolution for an initial condition $\rho_a=10^{10}$ GeV and $\delta=1$ and different values of $\kappa$.}
     \label{fig:caseradrad1e10}
\end{figure}

\begin{figure}[h!]
    \centering
     \subfigure[\label{fig:rr6}]{
         \centering
         \includegraphics[width=0.46\linewidth]{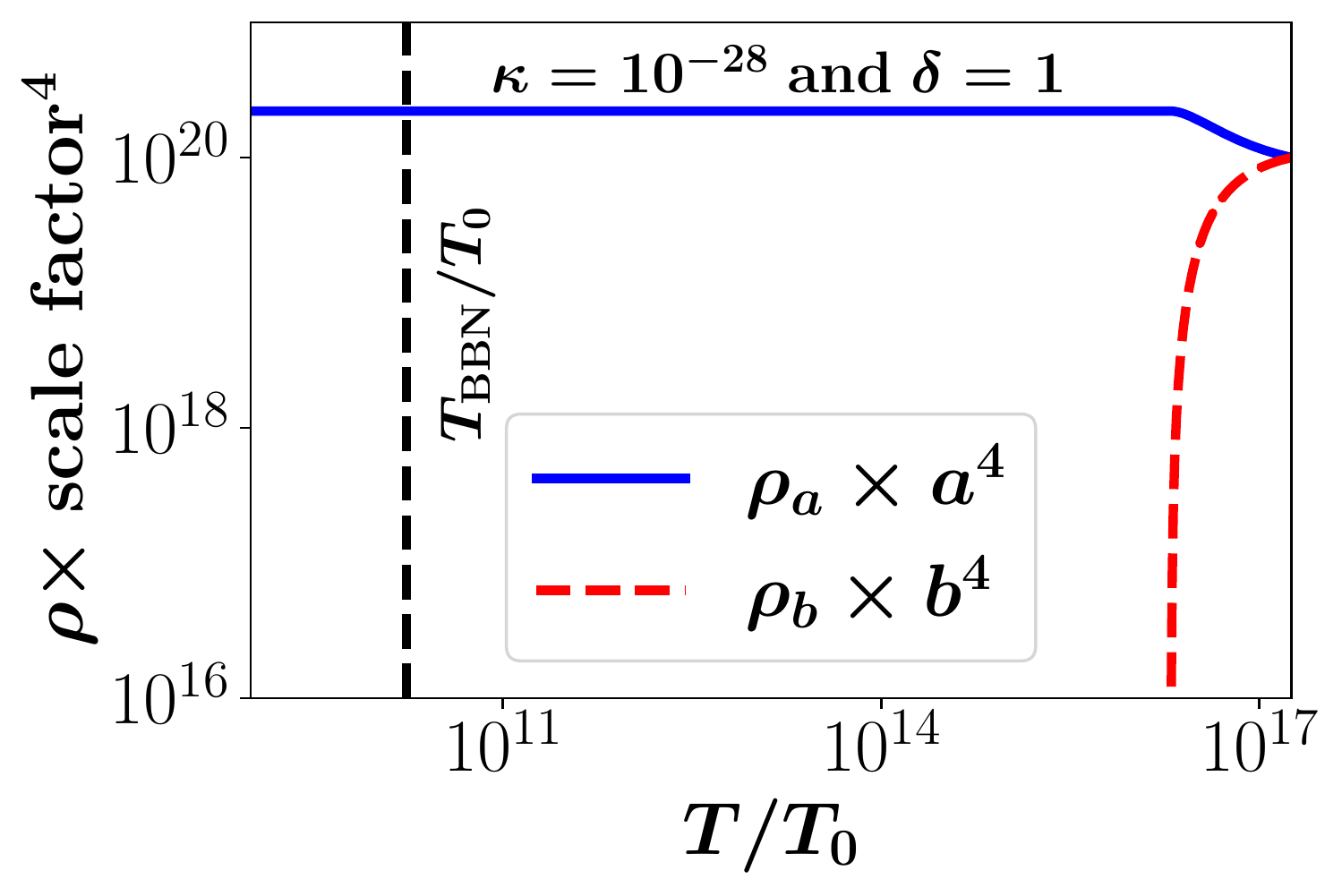}
     }
      \subfigure[\label{fig:rr8}]{
         \includegraphics[width=0.46\linewidth]{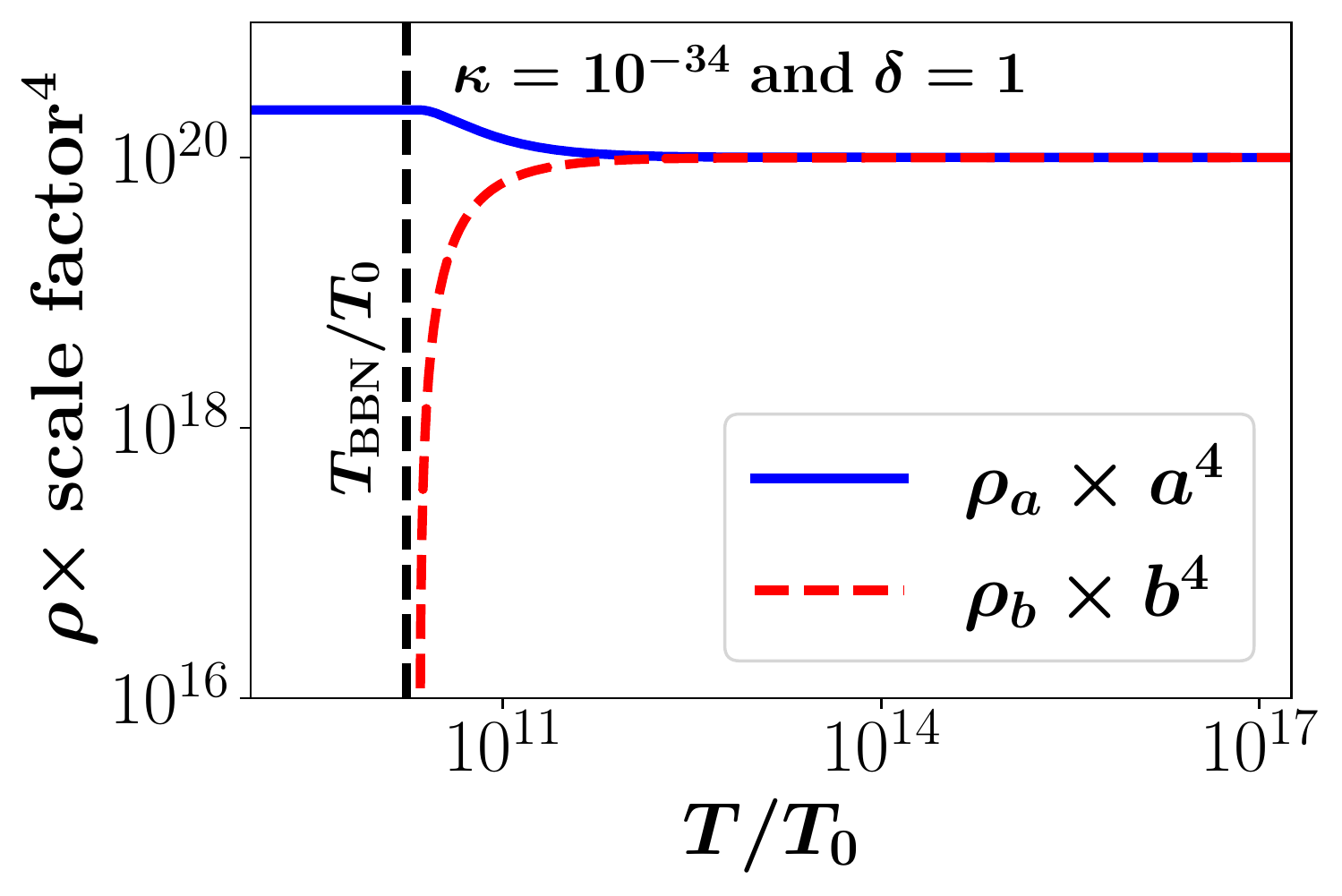}
     }
     \caption{Evolutution of radiation content in patch  $a$ and $b$ as function of the temperature. In the $y$ axes we have plotted $\rho_a\times a^4$ and     $\rho_b\times b^4$ . Panels (a) and (b) show the energy densities
     evolution for an initial condition $\rho_a=10^{20}$ GeV and $\delta=1$ an different values of $\kappa$}
     \label{fig:caseradrad1e20}
\end{figure}

\begin{figure}[h!]
    \centering
     \subfigure[\label{fig:rr10}]{
         \centering
         \includegraphics[width=0.46\linewidth]{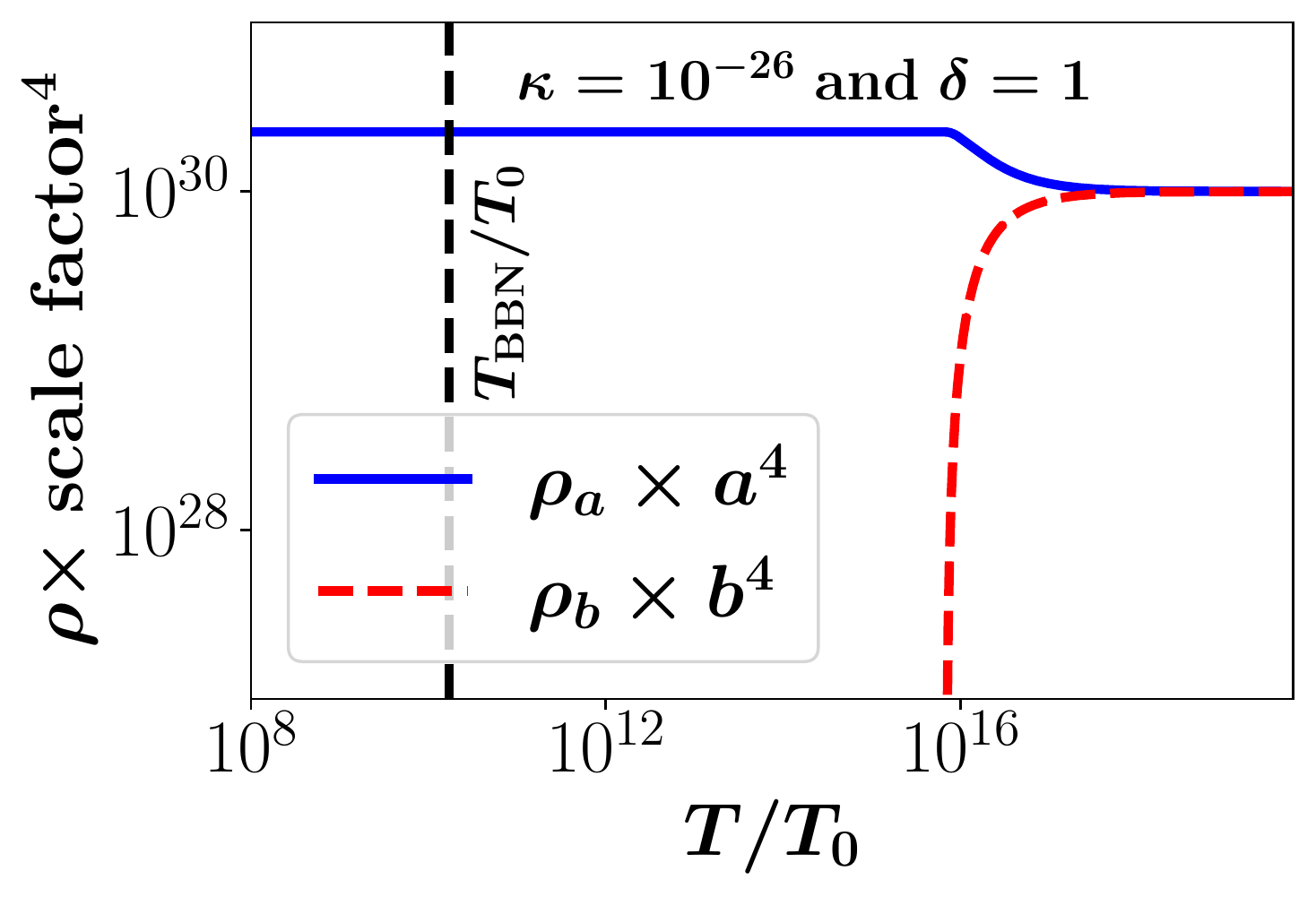}
     }
      \subfigure[\label{fig:rr12}]{
         \includegraphics[width=0.46\linewidth]{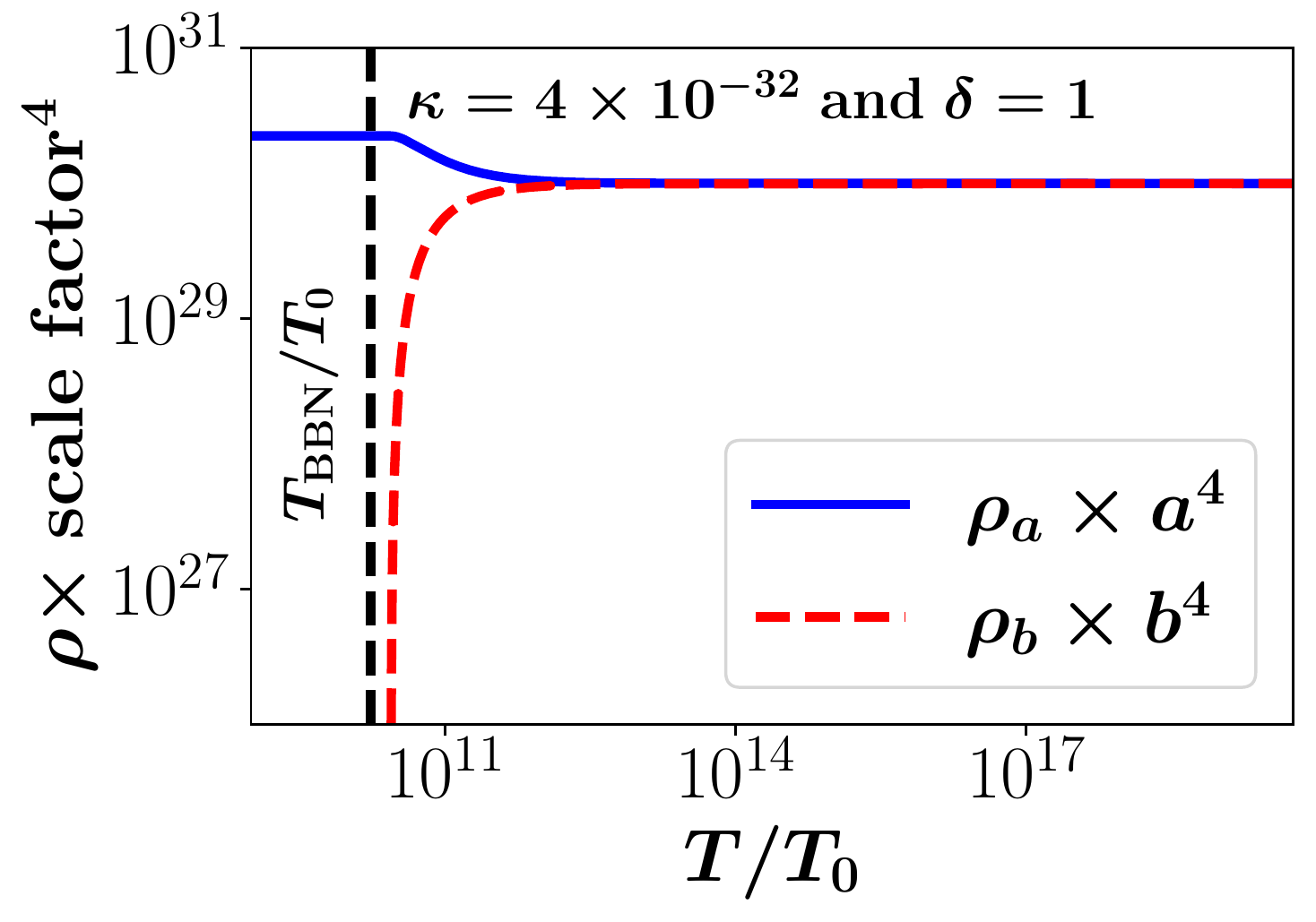}
     }
     \caption{Radiation content in patch  $a$ and $b$ as function of the temperature. Panels (a) and (b) show the energy densities
     evolution for an initial condition $\rho_a=10^{30}$ GeV and $\delta=1$.}
     \label{fig:caseradrad1e30}
\end{figure}

The evolution of energy densities as function of temperatures
is shown in Figures \ref{fig:caseradrad1e10} to \ref{fig:caseradrad1e30}
for different values of initial   $\rho_a$ and $\delta=1$.

We  observe for all cases how the  energy  density of relativistic matter in sector $a$ 
increases at expenses of the energy content of sector $b$ until the energy 
on this sector is completely drained. For the initial condition $\rho_a = 10^{10}$ GeV 
we can compare the relativistic--relativistic case depicted in 
Figure \ref{fig:caseradrad1e10} with the relatvistic--non-relativistic case in Figure 
\ref{fig:casematrad1e10}. Again, for large values of $\kappa$, the total
drain occurs for temperatures grater than the BBN temperature. The value 
of $\kappa$ at which the total  drain happens near $T_{BBN}$ is slightly smaller 
compared with the radiation-matter case.

The effects of a larger initial value of $\rho_a$ (with  $\delta =1$) are shown 
in Figures \ref{fig:caseradrad1e20} and \ref{fig:caseradrad1e30} which 
should be compared with Figures \ref{fig:casematrad1e20} (panels (b) and (c)) 
and \ref{fig:casematrad1e30}, respectively. The general features previously 
discussed are observed here and, additionally,  a small  value of  $\kappa$,  
 compared with the relativistic--non-relativistic case,  is necessary in order to reach  the total 
 drain at $T_{BBN}$.  In other words, for a fixed value of $\kappa$ and 
 initial $\delta=1$, the drain of energy from $b$ to $a$ happens faster
 if $b$ contains radiation compared with the case in which $b$ contains non-relativistic
 matter.

\section{Discussion and Conclusions}

In this work we have presented an extension of a cosmological model with two scale factors
in order to include matter. The two scale factors might represent two
sectors of a universe (two patches) \cite{RASOULI2019100269}, or even two different universes in a 
multiverse scenario \cite{Linde:2015edk}  which are causally connected only  through a
deformation  of the Poisson bracket structure. In this sense, this model is a sort of 
a non-commutative cosmology. The model is the analogous of the Landau problem in the 
space of metrics \cite{Falomir:2018ayx}.

The evolution of matter in such universe have been addressed and, in order to do that, 
we have assumed  a) the matter content on each  patch do not interact -- our 
matter-independent hypothesis -- and b) the modification of equations of motion is minimal
and it reduces to the usual equations of motion of General Relativity when the deformation
parameter $\kappa = 0$.  

Under such hypotheses, the equations of the evolution of the energy density have been 
solved numerically for two cases. In both, one of the patches contains relativistic matter
while the content of the other is relativistic in one case and non-relativistic in the second one.

The cases analyzed  shown an energy transfer from patch $b$ to patch $a$ in a sort of 
source-sink effect. The energy content  of $b$ drains completely to $a$ at some temperature
$T_{\mbox{\tiny{drain}}}$ which can be chosen to be equal to $T_{BBN}$ in order to restrict 
the possible values of the deformation parameter $\kappa$. Note that
the process is not symmetric under the change $a\leftrightarrow b$, since equations 
of motion do not have this symmetry. The system is symmetric under the previous change of scale factors
and $\kappa\rightarrow-\kappa$.

The rate at which the drain occurs (the function $\Gamma$ defined in (\ref{Gammaseq}))
depends on time through the scale factors and $\delta$ and it depends linearly on the 
deformation parameter $\kappa$.
In spite of this time (temperature) dependence, it is always possible to choose 
$\kappa$ so that the total drain happens at the desired temperature $T_{BBN}$ and 
this value of $\kappa$ will depend on the initial energy content of $a$ and $b$.

\begin{figure}[ht!]
    \centering
         \centering
         \includegraphics[width=0.6\linewidth]{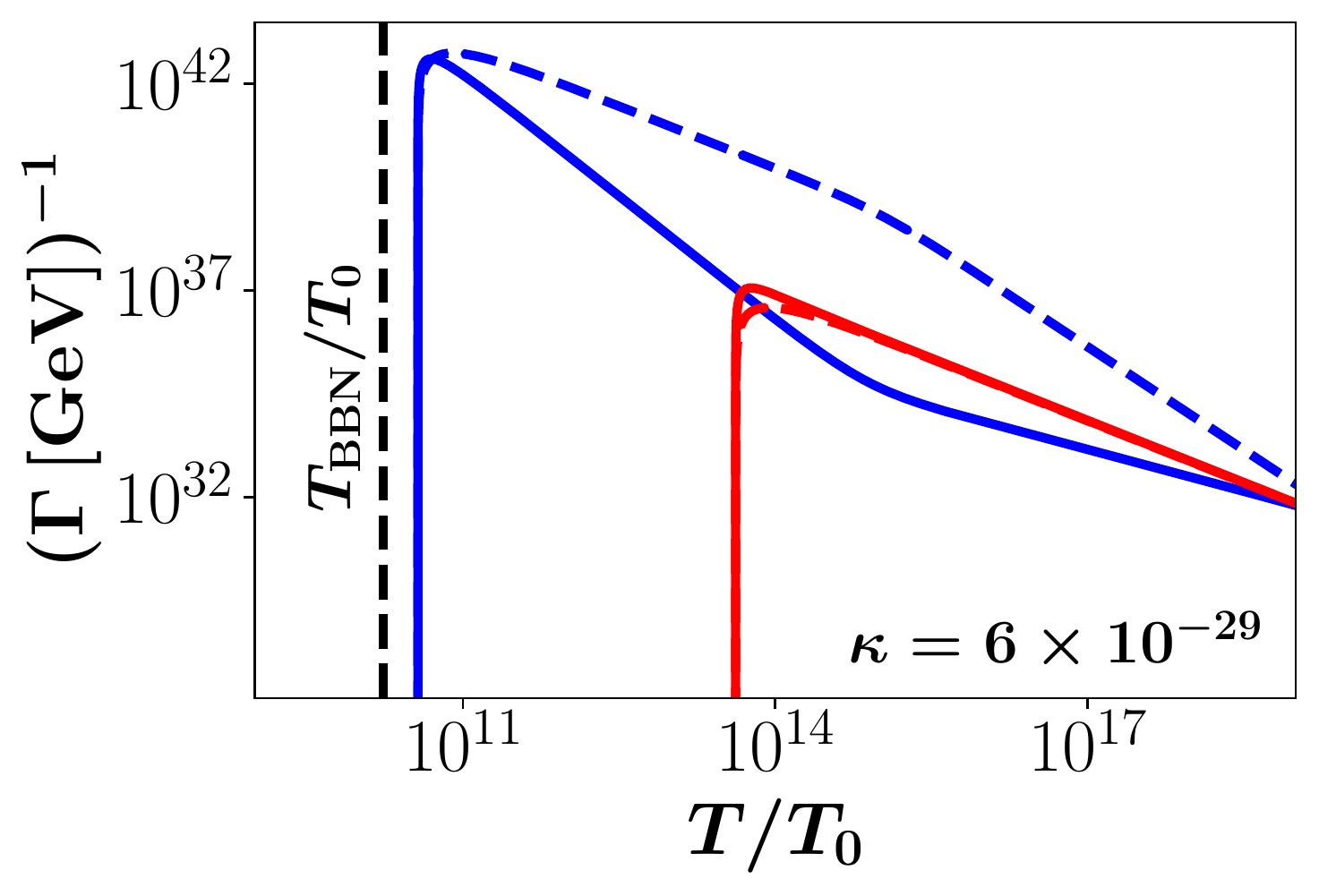}
     \caption{Evolution of  $\Gamma^{-1}$ in terms of the temperature. Solid lines represent $\Gamma_a$ and the dashed lines  $\Gamma_b$. In blue appears the  radiation-matter case (Sec. IV.A) while the   radiation-radiation (Sec.IV.B) is shown in red. It can be observed that for the same value of $\kappa$ the decay of radiation from $b$ to $a$ is faster than the case of matter. }
     \label{fig:Gamma}
\end{figure}

It is instructive  to compare the behaviour of the functions $\Gamma^{-1}$ 
(with dimensions of (time)$^{-1}$ )for  the radiation-radiation an radiation-matter cases.
In Figure \ref{fig:Gamma} -- where  blue lines correspond to  the radiation-matter  (section IV.A)
 and red lines to radiation-radiation   (section IV.B) -- this behaviour is shown. Here $\Gamma_a^{-1}$ appears in 
 solid line while dashed line is $\Gamma_b^{-1}$.

For fixed value of   $\kappa$  and initial $\delta =1$, we observe that   radiation 
decays faster than matter in patch $b$ or, in other words,   the energy drain of relativistic 
matter happens faster than the non-relativistic one.

Previous effect suggests  that the present model can be understood as  different type of non standard cosmology 
and it also suggests to include  dark matter in one of the patches. This analysis will be presented in forthcoming works.

\section{Acknowledgements}
This work was supported by Dicyt-USACH grants  USA1956-Dicyt (CM) and Dicyt-041931MF (FM).

\bibliography{biblio}

\end{document}